\pdfoutput=1

\newif\ifdraft\draftfalse  
\newif\ifsubmit\submitfalse 
\newif\iffull\fullfalse   
\newif\iflongrefs\longrefsfalse 
\newif\ifbackref\backreffalse 
\newif\ifsooner\soonerfalse
\newif\iflater\laterfalse
\newif\ifcamera\cameratrue    
\newif\ifcheckpagebudget\checkpagebudgetfalse
\newcommand{\xxx}{}
\makeatletter \@input{texdirectives.tex} \makeatother

\ifcamera
\documentclass[sigplan,screen]{acmart}


\else
\ifsubmit
\documentclass[sigplan,10pt,review]{acmart}
\else
\documentclass[sigplan,10pt]{acmart}
\fi
\fi


\copyrightyear{2018}
\acmYear{2018}
\ifcamera\setcopyright{acmlicensed}\else\setcopyright{rightsretained}\fi
\acmConference[CPP'18]{7th ACM SIGPLAN International Conference on Certified Programs and Proofs}{January 8--9, 2018}{Los Angeles, CA, USA}
\acmBooktitle{CPP '18: CPP '18: Certified Proofs and Programs , January 8--9, 2018, Los Angeles, CA, USA}
\acmPrice{15.00}
\acmDOI{10.1145/3167090}
\acmISBN{978-1-4503-5586-5/18/01}

\ifcamera\else
\makeatletter
\def\@copyrightpermission{This work is licensed under a \href{https://creativecommons.org/licenses/by/4.0/}{Creative Commons Attribution 4.0 International License}}
\makeatother
\fi

\ifcamera\else
\makeatletter
\def\@authorsaddresses{}
\fancypagestyle{firstpagestyle}{%
  \fancyhf{}%
  \renewcommand{\headrulewidth}{\z@}%
  \renewcommand{\footrulewidth}{\z@}%
    \fancyhead[L]{\ifsubmit\ACM@linecountL\fi}%
}
\fancypagestyle{standardpagestyle}{%
  \fancyhf{}%
  \renewcommand{\headrulewidth}{\z@}%
  \renewcommand{\footrulewidth}{\z@}%
    \fancyhead[LE]{\ifsubmit\ACM@linecountL\fi\@headfootfont\thepage}%
    \fancyhead[RO]{\@headfootfont\thepage}%
    \fancyhead[RE]{\@headfootfont\@shortauthors}%
    \fancyhead[LO]{\ifsubmit\ACM@linecountL\fi\@headfootfont\shorttitle}%
    \fancyfoot[RO,LE]{}%
}
\def\@mkbibcitation{}
\makeatother
\fi

\usepackage[justification=centering]{caption}
\usepackage{afterpage}
\usepackage{color}
\usepackage{amssymb,amsmath,amsfonts}
\usepackage{extarrows}
\usepackage{version}
\usepackage{xspace}
\usepackage{graphicx}
\usepackage{listings}
\usepackage{wrapfig}
\usepackage{ifpdf}
\usepackage{semantic}
\usepackage{mathpartir} 
\usepackage{natbib}
\setcitestyle{numbers}
\newcommand\citepos[1]{\citeauthor{#1}'s\ \citeyear{#1}}
\bibpunct();A{},
\usepackage{amsthm}
\usepackage{stmaryrd}
\usepackage{subfigure}
\usepackage{multirow}
\usepackage{proof}
\ifcamera
\usepackage{balance} 
\fi
\definecolor{darkblue}{rgb}{0.0,0.0,0.3}

\usepackage{hyperref}
\hypersetup{breaklinks=true}

\newcommand\maybecolor[1]{\color{#1}}

\newcommand\mypara[1]{\vskip 0.03in \noindent{\textbf{\itshape{#1}}}\quad}

\let\ls\lstinline

\def\Snospace~{\S{}}

\newcommand\fstar{F$^\star$\xspace}

\newcommand{\nat}{\mathbb{N}}

\newcommand{\Rand}[2]{#1 \stackrel{\raisebox{-.25ex}[.25ex]%
   {\tiny $\mathdollar$}}{\raisebox{-.2ex}[.2ex]{$\leftarrow$}} #2}
\newcommand{\zq}{\mathbb{Z}_q}

\definecolor{dkblue}{rgb}{0,0.1,0.5}
\definecolor{dkgreen}{rgb}{0,0.4,0}
\definecolor{dkred}{rgb}{0.6,0,0}
\definecolor{dkpurple}{rgb}{0.7,0,1.0}
\definecolor{purple}{rgb}{0.9,0,1.0}
\definecolor{olive}{rgb}{0.4, 0.4, 0.0}
\definecolor{teal}{rgb}{0.0,0.4,0.4}
\definecolor{azure}{rgb}{0.0, 0.5, 1.0}

\newcommand{\comm}[3]{\ifcheckpagebudget\else\ifdraft{\maybecolor{#1}[#2: #3]}\fi\fi}
\newcommand{\nik}[1]{\comm{dkpurple}{Nik}{#1}}
\newcommand{\ch}[1]{\comm{teal}{CH}{#1}}

\newcommand{\km}[1]{\comm{purple}{KM}{#1}}

\newcommand*{\EG}{e.g.,\xspace}
\newcommand*{\IE}{i.e.,\xspace}
\newcommand*{\ETAL}{et al.\xspace}

\newcommand\surl[1]{{\small\url{#1}}}
\newcommand\ssurl[1]{{\scriptsize\url{#1}}}

\begin{document}

\title{A Monadic Framework for Relational Verification}


\subtitle{Applied to Information Security, Program Equivalence, and Optimizations}

\xxx{}

\ifcamera
\author{Niklas Grimm} 
\affiliation{\institution{Vienna University of Technology}\country{Austria}}
\author{Kenji Maillard} 
\affiliation{\institution{Inria Paris and ENS Paris}\country{France}}
\author{C\'edric Fournet} 
\affiliation{\institution{Microsoft Research}\country{United Kingdom}}
\author{C\u{a}t\u{a}lin Hri\c{t}cu} 
\affiliation{\institution{Inria Paris}\country{France}}
\author{Matteo Maffei} 
\affiliation{\institution{Vienna University of Technology}\country{Austria}}
\author{Jonathan Protzenko} 
\affiliation{\institution{Microsoft Research}\country{United States}}
\author{Tahina Ramananandro} 
\affiliation{\institution{Microsoft Research}\country{United States}}
\author{Aseem Rastogi} 
\affiliation{\institution{Microsoft Research}\country{India}}
\author{Nikhil Swamy} 
\affiliation{\institution{Microsoft Research}\country{United States}}
\author{Santiago Zanella-B\'eguelin} 
\affiliation{\institution{Microsoft Research}\country{United Kingdom}}
\else
\author{\normalsize
      Niklas Grimm$^1$ \quad
      Kenji Maillard$^{2,3}$ \quad
      C\'edric Fournet$^4$ \quad
      C\u{a}t\u{a}lin Hri\c{t}cu$^2$ \quad
      Matteo Maffei$^1$ \quad
      Jonathan Protzenko$^4$ \quad
      Tahina Ramananandro$^4$ \quad
      Aseem Rastogi$^4$ \quad
      Nikhil Swamy$^4$ \quad
      Santiago Zanella-B\'eguelin$^4$\vspace{0.5em}}
\affiliation{$^1$Vienna University of Technology \qquad
            $^2$Inria Paris \qquad
            $^3$ENS Paris \qquad
            $^4$Microsoft Research\vspace{0.5em}}
\fi
\makeatletter
\renewcommand{\@shortauthors}{Grimm~\ETAL}
\makeatother

\begin{abstract}
Relational properties describe multiple runs of one or more programs.
They characterize many useful notions of security, program refinement,
and equivalence for programs with diverse computational effects, and
they have received much attention in the recent literature.
Rather than developing separate tools for special classes of effects and
relational properties, we advocate using a general purpose proof
assistant as a unifying framework for the relational verification of
effectful programs.
The essence of our approach is to model effectful computations using
monads and to prove relational properties on their monadic
representations, making the most of existing support for reasoning
about pure programs.

We apply this method in \fstar and evaluate it by encoding a variety of
relational program analyses, including information
flow control\iffull, semantic declassification\fi, program equivalence and
refinement at higher order, correctness of program
optimizations and game-based cryptographic security. 
By relying on SMT-based automation, unary weakest preconditions,
user-defined effects, and monadic reification, we show that,
compared to unary properties, verifying relational properties requires
little additional effort from the \fstar{} programmer.
\end{abstract}

\maketitle

\section{Introduction}
\label{sec:intro}

Generalizing unary properties (which describe single runs of programs),
\emph{relational} properties describe multiple runs of one or
more programs.
Relational properties are useful when reasoning about program
refinement, approximation, equivalence, provenance,
as well as many notions of security.
A great many relational program analyses have been
proposed in the recent literature, including works by \citet{Yang07,
ZaksP08, BentonKBH09, KunduTL09, GodlinS10, BartheGB12, BartheKOB13,
BartheFGSSB14, BartheGAHRS15, HedinS12, KustersTBBKM15, BanerjeeNN16,
AsadaSK16, CiobacaLRR16, AntonopoulosGHKTW17,BeckertKU15%
\iffull,BeckertBGHLU17\fi};
\citet{MurrayMBGBSLGK13};
\citet{FehrenbachC16};
\citet{BauereissG0R16,BauereissG0R17};
and \citet{CicekBG0H17}.
 While some systems have been designed for the efficient verification
 of specialized relational properties of programs (notably
 information-flow type systems, e.g., \citet{SabelfeldM03}), others
 support larger classes of properties.
These include tools based on product program constructions for
automatically proving relations between first-order imperative
programs (e.g., SymDiff \citep{LahiriHKR12} and Descartes
\citep{SousaD16}), as well as 
relational
program logics \cite{benton04relational} that support interactive
verification of relational properties within proof assistants~(e.g.,
EasyCrypt \citep{BartheGB12} and RHTT \citep{NanevskiBG13}).


We provide a framework in which relational logics and other
special-purpose tools can be recast on top of a general method for
relational reasoning.
%
The method is simple: we use monads to model and program effectful
computations; and we reveal the pure monadic representation of an
effect in support of specification and proof.
%
%
%
Hence, we reduce the problem of relating effectful computations to
relating their pure representations, and then apply the tools
available for reasoning about pure programs.
While this method should be usable for a variety of proof assistants,
we choose to work in \fstar \cite{mumon}, a dependently typed
programming language and proof assistant.
By relying on its support for SMT-based automation, unary
weakest preconditions, and user-defined effects \cite{dm4free}, we
demonstrate, through a diverse set of examples, that our approach enables
the effective verification of relational properties
with an effort comparable to proofs
of unary properties in \fstar and to proofs in relational logics with
SMT-based automation.

Being based on an expressive semantic foundation, our approach can be
directly used to verify relational properties of programs.
Additionally, we can still benefit from more specialized automated
proof procedures, such as syntax-directed relational type systems, by
encoding them within our framework.
Hence, our approach facilitates comparing and composing
special-purpose relational analyses with more general-purpose
semi-interactive proofs; and it encourages prototyping and
experimenting with special-purpose analyses with a path towards their
certified implementations.

\subsection{\iffull Relational reasoning via monadic reification:\fi{}
  A first example}
\label{sec:first-example}

We sketch the main \iffull elements of our method \else ideas \fi
on a proof of equivalence for
the two stateful, recursive functions below, a task not easily
accomplished using specialized relational program logics:

\begin{lstlisting}
let rec sum_up r lo hi = 
  if lo$\neq$hi then (r := !r+lo; sum_up r (lo+1) hi)
let rec sum_dn r lo hi = 
  if lo$\neq$hi then (r := !r+hi$-$1; sum_dn r lo (hi$-$1))
\end{lstlisting}
Both functions sum all numbers between \ls$lo$ and \ls$hi$ into some
accumulator reference \ls$r$, the former function by counting up and the latter
function by counting down.

\mypara{Unary reasoning about monadic computations}
As a first step, we embed these computations within a
dependently typed language. There are many proposals for how to
do this---one straightforward approach is to encapsulate effectful
computations within a parameterized monad \cite{atkey09parameterised}.
In \fstar, as in the original Hoare Type Theory \citep{nmb08htt}, these monads are
indexed by a computation's pre- and postconditions and proofs are
conducted using a unary program logic (\IE not relational), adapted for use with
higher-order, dependently typed programs.
Beyond state, \fstar supports reasoning about unary properties of a
wide class of user-defined monadic effects, where the monad can be
chosen to best suit the intended style of unary proof.

\mypara{Relating reified effectful terms} Our goal is
to conveniently state and prove properties that relate
effectful terms, e.g., prove \ls$sum_up$ and \ls$sum_dn$
equivalent.
We do so by revealing the monadic representation of these two
computations as pure state-passing functions.
However, since doing this na\"ively would preclude the efficient
implementation of primitive effects, such as state in terms of a
primitive heap, our general method relies on an explicit {\em monadic
reification} coercion for exposing the pure monadic representation of
an effectful computation in support of relational reasoning.%
\footnote{While this coercion is inspired by \citepos{Filinski94}
  \ls$reify$ operator, we only use it to reveal the pure
  representation of an effectful computation in support of
  specification and proof, whereas Filinski's main use of reification
  was to uniformly implement monads using continuations.}
Thus, in order to relate effectful terms, one simply reasons about
their pure reifications.
Turning to our example, we prove the following lemma,
stating that running \ls$sum_up$
and \ls$sum_dn$ in the same initial states produces equivalent final
states. (A proof is given in \S\ref{sec:lemmas}.)

\begin{lstlisting}
r:ref int -> lo:int -> hi:int{hi >= lo} -> h:heap{r $\in$ h} -> 
  reify (sum_up r lo hi) h $\sim$ reify (sum_dn r lo hi) h
\end{lstlisting}


\mypara{Flexible specification and proving style with SMT-\linebreak backed automation}
Although seemingly simple, proving \ls$sum_up$ and \ls$sum_dn$
equivalent is cumbersome, if at all possible, in most prior
relational program logics. Prior relational logics rely on common
syntactic structure and control flow between multiple programs to
facilitate the analysis. To reason about transformations like loop
reversal, rules exploiting syntactic similarity are not very useful
and instead a typical proof in prior systems may involve several
indirections, e.g., first proving the full functional correctness of
each loop with respect to a purely functional specification and then
showing that the two specifications are equivalent.
Through monadic reification, effectful terms are
\emph{self-specifying}, removing the need to rewrite the same
code in purely-functional style just to enable specification and reasoning.

Further, whereas many prior systems are specialized to proving binary
relations, it can be convenient to structure proofs using relations
of a higher arity, a style naturally supported by our method. For
example, a key lemma in the proof of the equivalence above is an
inductive proof of a ternary relation, which states
that \ls$sum_up$ is related to \ls$sum_up$ on a prefix combined
with \ls$sum_dn$ on a suffix of the interval \ls$[lo, hi)$.

Last but not least, using the combination of typechecking, weakest
precondition calculation, and SMT solving provided by \fstar, many
relational proofs go through with a degree of automation comparable to
existing proofs of unary properties,
as highlighted by the examples in this paper.

\subsection{Contributions and outline}



We propose a methodology for relational verification (\autoref{sec:ingredients}),
covering both broadly applicable ingredients such as representing
effects using monads and exposing their representation using monadic
reification, as well as our use of specific \fstar{} features that enable
proof flexibility and automation. All these ingredients are generic, \IE
none of them is specific to the verification of relational properties.

The rest of the paper is structured as a series of case studies
illustrating our methodology at work. Through these examples we aim to show
that our methodology enables comparing and composing various styles of
relational program verification in the same system, thus taking a step
towards unifying many prior strands of research.
Also these examples cover a wide range of applications that, when
taken together, exceed the ability of all previous tools for
relational verification of which we are aware.
Our examples are divided into three sections that can be read in any
order, each being an independent case study:

\mypara{Transformations of effectful programs (\autoref{sec:transformations})} 
We \linebreak develop an extensional, semantic characterization of a stateful program's
read and write effects, based on the relational approach
of \citet{benton06aplas}. Based on these semantic read and write effects, we derive
lemmas that we use to prove the correctness of common program
transformations, such as swapping the order of two commands and eliminating
redundant writes. Going further, we encode~\citepos{benton04relational}
relational Hoare logic in our system, providing a syntax-directed proof system
for relational properties as a special-purpose complement
to directly reasoning about a program's effects.
 
\mypara{Cryptographic security proofs (\autoref{sec:crypto})}
We show how to model basic game steps of code-based cryptographic
proofs of security \citep{Bellare2006} by proving equivalences between
probabilistic programs. We prove perfect secrecy of one-time pad
encryption \iffull, and a crucial lemma in the proof of semantic security of
ElGamal encryption\fi, an elementary use of \citepos{BartheGB09}
probabilistic relational Hoare logic.

\mypara{Information-flow control (\autoref{sec:ifc})} We encode
several styles of static information-flow control
analyses \iffull, while accounting for declassification\fi. Highlighting the
ability to compose various proof styles in a single framework, we
combine automated, type-based security analysis
with SMT-backed, semantic proofs of noninterference.

\mypara{Proofs of algorithmic optimizations (\autoref{sec:refinement})}
With a few exceptions, prior relational program logics apply to
first-order programs and provide incomplete proof rules that
exploit syntactic similarities between the related programs.
Not being bound by syntax, we prove relations of higher
arities (e.g., $4$-ary and $6$-ary relations) between higher-order,
effectful programs with differing control flow by reasoning directly
about their reifications. We present two larger
examples: First, we show how to memoize a recursive function
using \citepos{McBride15} partiality monad
and we prove it equivalent to the original non-memoized version.
Second, we implement an imperative union-find data structure, adding the
classic union-by-rank and path compression optimizations in several
steps and proving stepwise refinement.

\smallskip

From these case studies, we conclude that our method for relational
reasoning about reified monadic computations is both effective and
versatile.
We are encouraged to continue research in this direction, aiming to
place proofs of relational properties of effectful programs on an
equal footing with proofs of pure programs in \fstar as well as other
proof assistants and verification tools.

The code for the examples in this paper is available at\\
{\small \href{https://github.com/FStarLang/FStar/tree/master/examples/rel}{https://github.com/FStarLang/FStar/tree/master/examples/rel}}\\
Compared to this code, the listings in the paper are edited for
clarity and sometimes omit uninteresting details.
%
\ifcamera
The extended version \cite{relational} describes some
additional case studies that we omit here because of space.
\fi

\section{Methodology for relational verification}
\label{sec:ingredients}



In this section we review in more detail the key \fstar{}
features we use and how each of them contributes to our
verification method for relational properties.
Two of these features are general and broadly applicable:
(\autoref{sec:monads})~modeling effects using monads and keeping the
effect representation abstract to support efficient implementation of
primitive effects and
(\autoref{sec:reification})~using monadic reification to expose the
effect representation.
%
The remaining features are more specific to \fstar and
enable proof flexibility and automation:
(\autoref{sec:wp})~using a unary weakest precondition calculus to
  produce verification conditions in an expressive dependently typed
  logic;
(\autoref{sec:lemmas})~using dependent types together with pre- and
postconditions to express arbitrary relational properties of reified
computations;
(\autoref{sec:smt})~embedding the dependently typed logic into SMT
   logic to enable the SMT solver to reason by
   computation.

None of these generic ingredients is tailored to the verification
of relational properties, and while \fstar{} is currently the only
verification system to provide all these ingredients in a unified package,
each of them also appears in other systems.
This makes us hopeful that this relational verification method can
also be applied with other proof assistants (\EG Coq, Lean, Agda,
Idris, etc.), for which the automation would likely come in quite
different styles.



\subsection{Modeling effects using monads}
\label{sec:monads}

At the core of \fstar is a language of dependently typed, total
functions. Function types are written \ls$x:t -> Tot t'$ where
the co-domain \ls$t'$ may depend on the argument \ls$x:t$. Since it
is the default in \fstar{}, we often
drop the \ls$Tot$ annotation (except where needed for emphasis) and
also the name of the formal argument when it is unnecessary, e.g., we
write \ls$int -> bool$ for \ls$_:int -> Tot bool$. We also
write \ls$#x:t -> t'$ to indicate that the argument \ls$x$ is
implicitly instantiated.

Our first step is to describe effects using monads built
from total functions \cite{Moggi89}.
For instance, here is the standard monadic representation of state
in \fstar{} syntax.


%
%
%
%
\begin{lstlisting}
type st (mem:Type) (a:Type) = mem -> Tot (a * mem)
\end{lstlisting}
This defines a type \ls$st$ \ifsubmit indexed \else parameterized \fi
by types for the memory (\ls$mem$) and the result (\ls$a$).
We use \ls$st$  as the representation type of a new \ls$STATE_m$ effect we
add to \fstar, with the \ls$total$ qualifier enabling the termination
checker for \ls$STATE_m$ computations.
\begin{lstlisting}
total new_effect {$\label{state}$
 STATE_m (mem:Type) : a:Type -> Effect
 with repr = st mem;
  return = fun (a:Type) (x:a) (m:mem) -> x, m;
  bind = fun (a b:Type) (f:st mem a) (g:a -> st mem b) (m:mem) -> 
   let z, m' = f m in g z m';
  get = fun () (m:mem) -> m, m;  put = fun (m:mem) _ -> (), m  }
\end{lstlisting}
This defines the \ls$return$ and \ls$bind$ of this monad,
and two actions: \ls$get$ for obtaining the current memory,
and \ls$put$ for updating it.
The new effect \ls$STATE_m$ is still parameterized by the type of memories,
which allows us to choose a memory model best suited to the programming
and verification task at hand. We often instantiate \ls$mem$
to \ls$heap$ (a map from references to their values, as in ML), obtaining
the \ls$STATE$ effect shown below---we use other memory types
in \autoref{sec:ifc} and~\autoref{sec:equiv}.
\begin{lstlisting}
total new_effect STATE = STATE_m heap
\end{lstlisting}


While such monad definitions could in principle be used to directly
extend the implementation of any functional language with the state effect, a
practical language needs to allow keeping the representation of some
effects abstract so that they are efficiently implemented
primitively \cite{PeytonJones01}.
%
\fstar{} uses its simple module system to keep the monadic representation
of the \ls$STATE$ effect abstract and implements it under the hood
using the ML heap, rather than state passing (and similarly for
other primitive ML effects such as exceptions).
Whether implemented primitively or not, the monadic definition of each
effect is always the {\em model} used by \fstar{} to reason
about effectful code, both intrinsically using a (non-relational)
weakest precondition calculus (\autoref{sec:wp}) and extrinsically
using monadic reification (\autoref{sec:reification}).

For the purpose of verification, monads provide great flexibility in
the modeling of effects, which enables us to express relational
properties and to conduct proofs at the right level of abstraction.
For instance, \iffull in \autoref{sec:declassification} we extend a state
monad with extra ghost state to track declassification, \fi  in
\autoref{sec:crypto} we define a monad for random sampling from a
uniform distribution, and  in \autoref{sec:memo} we define a partiality
monad for memoizing recursive functions.
Moreover, since the difficulty of reasoning about effectful code is
proportional to the complexity of the effect, we do not use a single
full-featured monad for all code; instead we define custom monads for
sub-effects and relate them using monadic lifts.
For instance, we define a \ls$READER$ monad for computations that only
read the store, lifting \ls$READER$ to \ls$STATE$ only where
necessary (\autoref{sec:ifc-while} provides a detailed example).
While \fstar{} code is always written in an ML-like direct style, the
\fstar{} typechecker automatically inserts binds, returns and lifts under
the hood \cite{swamy11coco}.

\subsection{Unary weakest preconditions for user-defined effects and intrinsic proof}
\label{sec:wp}

For each user-defined effect, \fstar derives a weakest precondition
calculus for specifying unary properties and computing verification
conditions for programs using that effect \cite{dm4free}.
Each effect definition induces a computation type indexed by a
predicate transformer describing that computation's effectful
semantics.

For state, we obtain a computation type
`\ls$STATE a wp$' indexed by a result type \ls$a$ and by~\ls$wp$,
a predicate transformer of type
\ls$(a -> heap -> Type) -> heap -> Type$,
mapping postconditions (relating the result and final state of the
computation) to preconditions (predicates on the initial state).
\iffull For example, t\else{T}\fi{he} types of the \ls$get$ and \ls$put$ actions
of \ls$STATE$ are specified as:

\begin{lstlisting}
val get : unit -> STATE heap (fun post (h:heap) -> post h h)
val put : h':heap -> STATE unit (fun post (h:heap) -> post () h')
\end{lstlisting}

\noindent The type of \ls$get$ states that, in order to prove any
postcondition \ls$post$ of `\ls$get ()$' evaluated in state \ls$h$, it
suffices to prove \ls$post h h$, whereas for \ls$put h'$ it suffices
to prove \ls$post () h'$.
%
%
\fstar users find it more convenient to index computations with pre- and
postconditions as in HTT \cite{nmb08htt}, or sometimes
not at all, using the following abbreviations:

\begin{lstlisting}
ST a (requires p) (ensures q) = STATE a (fun post h$_0$ -> 
             p h$_0$ /\ (forall (x:a) (h$_1$:heap). q h$_0$ x h$_1$ ==> post x h$_1$))
St a = ST a (requires (fun _ -> True)) (ensures (fun _ _ _ -> True))
\end{lstlisting}

\fstar{} computes weakest preconditions
generically for any effect.
Intuitively, this works by putting the code into an explicit monadic
form and then translating the binds, returns, actions, and lifts from the
expression level to the weakest precondition level.
This enables a convenient form of \emph{intrinsic} proof in \fstar,
i.e., one annotates a term with a type capturing properties of
interest; \fstar computes a weakest precondition for the term and
compares it to the annotated type using a built-in subsumption rule,
checked by an SMT solver.

For example, \iffull in the code below, \fstar{} checks that the inferred
computation type is sufficient to prove that a \ls$noop$ function leaves the
memory unchanged.


For a more interesting example,\fi the \ls$sum_up$ function
from \S\ref{sec:first-example} can be given the following type:

\begin{lstlisting}
r:ref int -> lo:nat -> hi:nat{hi >= lo} -> 
          ST unit (requires fun h -> r $\in$ h) (ensures fun _ _ h -> r $\in$ h)
\end{lstlisting}

This is a dependent function type, for a function with three
arguments \ls$r$, \ls$lo$, and \ls$hi$ returning a terminating, stateful
computation.
The \emph{refinement} type \ls$hi:nat{hi >= lo}$ restricts
\ls$hi$ to only those natural numbers greater than or equal to \ls$lo$.
The computation type of `\ls$sum_up r lo hi$' simply requires and ensures
that its reference argument \ls$r$ is present in the memory.
\fstar computes a weakest precondition from the implementation of
\ls$sum_up$ (using the types of \ls$(!)$ and \ls$(:=)$ 
provided by
the
\ls$heap$ memory model used by \ls$STATE$) and proves that its
inferred specification is subsumed by the user-provided annotation.
The same type can also be given to \ls$sum_dn$.

\subsection{Exposing effect definitions via \iffull monadic \fi reification}
\label{sec:reification}

Intrinsic proofs of effectful programs in \fstar are inherently
restricted to unary properties. Notably, pre- and postconditions
are required to be pure terms, making it impossible for specifications
to refer directly to effectful code,
e.g., \ls$sum_up$ cannot directly use itself or \ls$sum_dn$ in its
specification. To overcome this restriction, we need a way to coerce a terminating
effectful computation to its underlying monadic representation
which is a pure term---\citepos{Filinski94} monadic
reification provides just that facility.%
\footnote{Less frequently, we use {\tt reify}'s dual, {\tt reflect},
  which packages a pure function as an effectful computation.
}


Each new effect in \fstar{} induces a \ls$reify$ operator
that exposes the representation of an effectful computation
in terms of its underlying monadic representation \cite{dm4free}.
For the \ls$STATE$ effect, \fstar provides the following (derived) rule
for \ls$reify$, to coerce a stateful computation to a total,
explicitly state-passing function of type \ls$heap -> t * heap$.
The argument and result types of
\ls$reify$$~e$ are refined to capture the pre- and postconditions
intrinsically proved for $e$.
\[\footnotesize
\inferrule{
  S;\Gamma |- e : \text{\ls$ST t (requires pre) (ensures post)$}
}
{
  S;\Gamma |- \text{\ls$reify$}~e:
              \text{\ls$h:heap\{pre h\} -> Tot (r:(t*heap)\{post h (fst r) (snd r)\})$}
}
\]

The semantics of \ls$reify$ is to traverse the term and
to gradually expose the underlying
monadic representation. We illustrate this below for \ls$STATE$,
where the constructs on the right-hand side of the rules
are the pure implementations
of \ls$return$, \ls$bind$, \ls$put$, and \ls$get$ as defined on
page~\pageref{state}, but with type arguments left implicit:
\newcommand{\steps}[0]{\ensuremath{\leadsto}}
\newcommand{\stepss}[0]{\ensuremath{\steps^{*}}}
\[\small
\begin{array}{r@{\hspace{0.5em}}c@{\hspace{0.5em}}l}
\text{\ls$reify (return $}~e\text{\ls$)$} &\steps& \text{\ls$STATE.return $}~e\\
\text{\ls$reify (bind x $}\leftarrow e_1~\text{\ls$ in $}~e_2\ls$)$ &\steps&
\begin{array}{l}
  \hspace{-5pt}\text{\ls$STATE.bind (reify $}~e_1\text{\ls$)(fun x->reify $}~e_2\text{\ls$)$}
\end{array} \\
\text{\ls$reify (get $}~e\text{\ls$)$} &\steps& \text{\ls$STATE.get $}~e\\
\text{\ls$reify (put $}~e\text{\ls$)$} &\steps& \text{\ls$STATE.put $}~e
\end{array}
\]

Armed with \ls$reify$, we can write an \emph{extrinsic} proof of a
lemma relating \ls$sum_up$ and \ls$sum_dn$ (discussed in detail
in \S\ref{sec:lemmas}), i.e., an ``after the fact'' proof that
is separate from the definition of \ls$sum_up$ and \ls$sum_dn$
and that relates their reified executions.
We further remark that in \fstar{} the standard operational semantics
of effectful computations is modeled in terms of reification, so
proving a property about a reified computation is really the same as
proving the property about the evaluation of the computation itself.

The \ls$reify$ operator
clearly breaks the abstraction of the underlying monad and needs to be used
with care. \citet{dm4free} show that programs that do not use \ls$reify$
(or its converse, \ls$reflect$)
can be compiled efficiently. Specifically, if the
computationally relevant part of a program is free of \ls$reify$
then the \ls$STATE$ computations can be compiled
using primitive state with destructive updates.

To retain these benefits of abstraction, we rely on \fstar's module
system to control how the abstraction-breaking \ls$reify$ coercion
can be used in client code.
In particular, when abstraction violations cannot be tolerated, we
use \fstar's \ls$Ghost$ effect (explained in
\S\ref{sec:lemmas}) to mark \ls$reify$
as being usable only in computationally irrelevant code, limiting the use
of monadic reification to specifications and proofs.
This allows one to use reification even though effects like state and
exceptions are implemented primitively in \fstar.

\subsection{Extrinsic specification and proof, eased by SMT-based automation}
\label{sec:lemmas}
\label{sec:smt}


%
We now look at the proof relating \ls$sum_up$ and \ls$sum_dn$ in
detail, explaining along the way several \fstar-specific idioms that
we find essential to making our method work well.

\mypara{Computational irrelevance (\ls$Ghost$ effect)}
The \ls$Ghost$ effect is used to track a
form of computational irrelevance.
\ls$Ghost t (requires pre) (ensures post)$ is the type of a
pure computation returning a value of type \ls$t$ satisfying
\ls$post$, provided \ls$pre$ is valid.
However, this computation must be erased before running the program,
so it can only be used in specifications and proofs.

\mypara{Adding proof irrelevance (\ls$Lemma$)} 
\fstar
provides two closely related forms of proof irrelevance. First, a pure
term \ls$e:t$ can be given the refinement type \ls|x:t{$\phi$}| when it
validates the formula \ls@$\phi$[e/x]@, although no proof of $\phi$ is
materialized. For example, borrowing the terminology
of \citet{Nogin02}, the value \ls$()$ is a \emph{squashed} proof
of \ls@u:unit{0 <= 1}@. Combining proof and computation irrelevance,
\ls$e : Ghost unit pre (fun () -> post)$ is a squashed proof of \ls$pre -> post$.
This latter form is so common that we write it as
\ls$Lemma (requires pre) (ensures post)$,
further abbreviated as \ls$Lemma post$ when \ls$pre$ is \ls$True$.

\mypara{Proof relating \ls$sum_up$ and \ls$sum_dn$}
Spelling out the main lemma of \S\ref{sec:first-example}, our goal is
a value of the following type:

\begin{lstlisting}
val eq_sum_up_dn (r:ref int)(lo:int)(hi:int{hi >= lo})(h:heap{r $\in$ h})
: Lemma 
  (v r (reify (sum_up r lo hi) h) == v r (reify (sum_dn r lo hi) h))
\end{lstlisting}
where \ls$v r (_, h) = h.[r]$ and \ls$h.[r]$ selects the contents of the
reference \ls$r$ from the heap \ls$h$.

An attempt to give a trivial definition
for \ls$eqsum_up_dn$ that simply returns a unit value
\ls$()$ fails, because the SMT solver
cannot automatically prove the strong postcondition above.
Instead our proof involves calling an auxiliary
lemma \ls$sum_up_dn_aux$, proving a ternary relation:
\begin{lstlisting}
val sum_up_dn_aux (r:ref int) (lo:int) (mid:int{mid >= lo}) 
                   (hi:int{hi >= mid}) (h:heap{r $\in$ h})
: Lemma  (v r (reify (sum_up r lo hi) h)
         == v r (reify (sum_dn r lo mid) h) 
           + v r (reify (sum_up r mid hi) h) $-$ h.[r])
  (decreases (mid $-$ lo))
let eq_sum_up_dn r lo hi h = sum_up_dn_aux r lo hi hi h
\end{lstlisting}
While the statement of \ls$eq_sum_up_dn$ is different from the
statement of \ls$sum_up_dn_aux$, the SMT-based automation fills in the
gaps and accepts the proof sketch.
In particular, the SMT solver figures out that \ls$sum_up r hi hi$ is
a no-op by looking at its reified definition.
In other cases, the user has to provide more interesting proof sketches
that include not only calls to lemmas that the SMT solver cannot
automatically apply but also the cases of the proof and the recursive
structure.  This is illustrated by the \iffull\else following \fi
proof\iffull of \ls$sum_up_dn_aux$\fi:
\begin{lstlisting}
let rec sum_up_dn_aux r lo mid hi h =
  if lo $\neq$ mid then (sum_up_dn_aux r lo (mid $-$ 1) hi h;
                  sum_up_commute r mid hi (mid $-$ 1) h;
                  sum_dn_commute r lo (mid $-$ 1) (mid $-$ 1) h)
\end{lstlisting}
This proof is by induction on the difference between \ls$mid$ and
\ls$lo$ (as illustrated by the \ls$decreases$ clause of the lemma,
this is needed because we are working with potentially-negative integers). If
this difference is zero, then the property is trivial since the SMT
solver can figure out that \ls$sum_dn r lo lo$ is a no-op.
%
Otherwise, we call \ls$sum_up_dn_aux$ recursively for \ls+mid $-$ 1+ as
well as two further commutation lemmas (not shown) about \ls$sum_up$
and \ls$sum_dn$ and the SMT automation can take care of the rest.




\mypara{Encoding computations to SMT}
%
So how did \fstar{} figure out automatically that \ls$sum_up r hi hi$
and \ls$sum_dn r lo lo$ are no-ops?
For a start the \fstar{} normalizer applied the semantics of
\ls$reify$ sketched in \autoref{sec:reification} to partially evaluate the
term and reveal the monadic
representation of the \ls$STATE$ effect by traversing the term and
unfolding the monadic definitions of return, bind, actions and lifts.
In the case of \ls$reify (sum_up r hi hi) h$, for instance, reduction
intuitively proceeds as follows:
\begin{lstlisting}
reify (sum_up r hi hi) h 
  $\steps$ reify (if hi $\neq$ hi then (r := !r + lo; sum_up r (lo + 1) hi)) h
  $\stepss$ if hi $\neq$ hi then (STATE.bind (reify (Ref.read r) h) (fun x ->
                    STATE.bind (reify (Ref.upd r (x + lo))) 
                      (fun _ -> reified_sum_up r (hi + 1) hi))) h
      else STATE.return () h
  $\stepss$ if hi $\neq$ hi then let x, h' = reify (Ref.read r) h in
                    let _, h'' = reify (Ref.upd r (x + lo)) h' in
                    reified_sum_up r (hi + 1) hi h''
      else ((), h)
\end{lstlisting}

What is left is pure monadic code that \fstar{} then encodes to
the SMT solver in a way that allows it to reason by
computation \cite{AlejandroHKS16}.
For \ls$reify (sum_up r hi hi) h$ the SMT solver can
trivially show that \ls|hi $\neq$ hi| is false and thus the
computation returns the pair \ls$((), h)$.

While our work did not require any extension to \fstar{}'s
theory~\cite{dm4free}, we significantly improved \fstar{}'s logical
encoding to perform normalization of open terms based on the semantics
of reify (a kind of symbolic execution) before calling the SMT
solver. This allowed us to scale and validate the theory of
\citet{dm4free} from a single 2-line example to the $\approx$4,300
lines of relationally verified code presented in this paper.

\subsection{Empirical evaluation of our methodology}

For this first example, we reasoned directly about the semantics of two
effectful terms to prove their equivalence.
However, we often prefer more structured reasoning principles to prove
or enforce relational properties, e.g., by using program logics,
syntax-directed type systems, or even dynamic analyses.
In the rest of this paper, we show through several case studies, that
these approaches can be accommodated, and even composed, within our
framework.

Table~\ref{fig:perf} summarizes the empirical evaluation from these
case studies.
Each row describes a specific case study, its size in lines of source code,
and the verification time using \fstar and the Z3-4.5.1 SMT
solver. The verification times were collected on an Intel Xeon E5-2620
at 2.10 GHz and 32GB of RAM.
The ``1st run'' column indicates the time it takes \fstar and Z3
to find a proof. This proof is then used to generate hints (
unsat cores) that can be used as a starting point to verify subsequent
versions of the program.
The ``replay'' column indicates the time it takes to verify the
program given the hints recorded in the first run.
Proof replay is usually significantly faster, indicating that although
finding a proof may initially be quite expensive, revising a proof
with hints is fast, which greatly aids interactive proof development.

\begin{table}[t]
  \small
  
\begin{tabular}{|l|r|r|r|r|}
\hline
Subject&Section&1st run (ms)&Replay (ms)&Loc
\\\hline
Loops&{\ref{sec:first-example}}&218192&8943&127\\
Reorderings&{\ref{sec:transformations-spec}}&9239&4749&158\\
Benton (2004)&{\ref{sec:benton2004-rhl}}&832706&22920&1352\\
Cryptography&{\ref{sec:crypto}}&17307&10015&530\\
Static IFC&{\ref{sec:ifc-while}}&68525&15909&730\\
Hybrid IFC&{\ref{sec:ifc-combining}}&55472&1038&34\\
Declassification&\ifcamera{*}\else\ref{sec:declassification}\fi&63763&9811&208\\
IFC Monitor&\ifcamera{*}\else\ref{sec:ifc-monitor}\fi&44589&11480&502\\
Memoization&{\ref{sec:memo}}&12198&12294&427\\
Union-find&{\ref{sec:unionfind}}&89838&33455&295\\
\hline Total&&1411829&130614&4363\\\hline
\end{tabular}

  \caption{Code size (lines of code without comments) and proof-checking time (ms) for our examples. \ifcamera Examples with label * appear in the extended version~\cite{relational}.\fi}
\label{fig:perf}
\ch{Didn't want to touch the generated file (too much), but I would change ref
  with autoref and add subsection 2.4 for Loops. I would also put the
  Loc column before the timings. Finally, I would round and switch to
  seconds, since I don't think our numbers are precise all the way to
  milliseconds; and the numbers look much too big in milliseconds.}
\end{table}

\section{Correctness of program transformations}
\label{sec:transformations}

\nik{About section 3 in the paper (program transformations): a missed opportunity ... I realize now that we should have explicitly made several connections there
1. Some of the equivalences we proved have a very single-sided flavor, the last one in particular. We should have said so, making connections to RHL (edited)
2. We're very close in s3 to have a shallowly embedded while language (we have seq and conditionals, we should have added while)
and we should have pointed out the shallow/deep contrast w.r.t the IFC section
It would be nice to derive all the rules in some canonical RHL for while on this shallowly embedded language ... we're not very far from it}

Several researchers have devised custom program logics for verifying
transformations of imperative
programs \citep{benton04relational, BartheGB09, CarbinKMR12}.
We show how to derive similar rules justifying the correctness of
generic program transformations within our monadic framework.
We focus on stateful programs with a fixed-domain, finite memory.
We leave proving transformations of commands that dynamically allocate
memory to future work.\ch{``in a somewhat different context,
  \S\ref{sec:refinement} contains examples that use dynamic allocation
  and local state'' This last claim seems wrong; paper doesn't show
  anything about this, and not sure if it's true even in the
  implementation}

\let\li\ls


\subsection{Generic transformations based on read- and write-footprints}
\label{sec:transformations-spec}

Here and in the next subsection,
we represent a command $c$ as a function of type \ls$unit -> St unit$
that may read or write arbitrary references in memory.
\begin{lstlisting}
type command = unit -> St unit
\end{lstlisting}
In trying to validate transformations of commands, it is traditional to
employ an effect system to delimit the parts of memory that a
command may read or write.
Most effect systems are unary, syntactic analyses. For example,
consider the classic frame rule from separation logic:
\[
\{ P \} c \{ Q \} \Rightarrow
\{ P \ast R \} c \{ Q \ast R \}
\]
The command $c$ requires ownership of a subset of the heap $P$ in
order to execute, then returns ownership of $Q$ to its
caller. Any distinct heap fragment $R$ remains unaffected by the
function. Reading this rule as an effect analysis, one may conclude
that $c$ may read or write the $P$-fragment of memory---however, this
is just an approximation of $c$'s extensional behavior.
\citet{benton06aplas} observe that a more precise, semantic characterization
of effects arises from a relational perspective. Adopting this
perspective, one can define the \ls$footprint$ of a command
extensionally, using two unary properties and one binary property.

Capturing a command's write effect is easy with a
unary property, `\ls$writes c ws$' stating that the initial and final
heaps agree on the contents of their references, except \iffull for \fi those
in \iffull the set\fi \ls$ws$.
\begin{lstlisting}
type addrs = S.set addr
let writes (c:command) (ws:addrs) = forall (h:heap).
 let h' = snd  (reify (c ()) h) in
 (forall r. r $\in$ h <==> r $\in$ h') /\ (* no allocation *)
 (forall r. addr_of r $\not\in$ ws ==> h.[r] == h'.[r]) (* no changes except ws*)
\end{lstlisting}

Stating that a command only reads references \ls$rs$ is similar in
spirit to \iffull the statement of \fi noninterference (\iffull
to which we return in \fi\S\ref{sec:ifc-while}).
Interestingly, it is impossible to describe
the set of locations that a command may read without also speaking
about the locations it may write. The relation `\ls$reads c rs ws$'
states that if \ls$c$ writes at most the references in \ls$ws$, then
executing \ls$c$ in heaps that agree on the references in \ls$rs$
produces heaps that agree on \ls$ws$, i.e., \ls$c$ does not depend on
references outside \ls$rs$.

\ch{While the comment above helps intuition a bit, the writes and
  reads names really don't. Guido, Kenji, and me were all confused.
  not sure about the others, but the reason I was confused was that I
  was expecting writes to be the only thing that restricts writes and
  reads to only restrict reads}

\ch{Santiago: I realized this is exactly what we called Observational
  Equivalence in CertiCrypt (section 4.2 in
  \url{https://www.microsoft.com/en-us/research/wp-content/uploads/2016/02/Zanella.2009.POPL_.pdf})
  One funny thing is that if you have \ls$c:cmd rs ws$, then \ls$c:cmd rs (S.union rs ws)$ also holds.}

\begin{lstlisting}
let equiv_on (rs:addr_set) (h$_0$:heap) (h$_1$:heap) =
 forall a (r:ref a). addr_of r $\in$ rs /\ r $\in$ h$_0$ /\ r $\in$ h$_1$ ==> h$_0$.[r] == h$_1$.[r]
let reads (c:command) (rs ws:addrs) = forall (h$_0$ h$_1$: heap).
 let h'$_0$, h'$_1$ = snd (reify (c ()) h$_0$), snd (reify (c ()) h$_1$) in
 (equiv_on rs h$_0$ h$_1$ /\ writes c ws) ==> equiv_on ws h'$_0$ h'$_1$
\end{lstlisting}

Putting the pieces together, we define a read- and
write-footprint-indexed type for commands:

\begin{lstlisting}
type cmd (rs ws:addrs) = c:command{writes c ws /\ reads c rs ws}
\end{lstlisting}

\newcommand\bindseq{\ensuremath{\mbox{\texttt{>>}}}}

One can also define combinators to manipulate footprint-indexed
commands. For example, here is a `\bindseq' combinator for sequential
composition. Its type proves that read and write-footprints compose by
a pointwise union, a higher-order relational property; the proof
requires an (omitted) auxiliary lemma \ls$seq_lem$ (recall that
variables preceded by a \ls$#$ are implicit arguments):

\begin{lstlisting}
let seq (#r1 #w1 #r2 #w2 : addrs) (c1:cmd r1 w1) (c2:cmd r2 w2) : 
  command = c1(); c2()
let ($\bindseq$) #r1 #w1 #r2 #w2 (c1:cmd r1 w1) (c2:cmd r2 w2) :
  cmd (r1 $\cup$ r2) (w1 $\cup$ w2) = seq_lem c1 c2; seq c1 c2
\end{lstlisting}

\iffull
\subsection{Several transformations on commands}
\label{sec:transformations-proofs}
\fi

Making use of relational footprints, we can prove other relations
between commands, e.g., equivalences that justify program
transformations. Command equivalence \ls@c$_0$ $\sim$ c$_1$@ states
that running \ls@c$_0$@ and \ls@c$_1$@ in identical initial heaps
produces (extensionally) equal final heaps.

\begin{lstlisting}
let ($\sim$) (c$_0$:command) (c$_1$:command) = forall h.
 let h$_0$, h$_1$ = snd (reify (c$_0$ ()) h), snd (reify (c$_1$ ()) h) in
 forall (r:ref 'a). (r $\in$ h$_0$ <==> r $\in$ h$_1$) /\ (r $\in$ h$_0$ ==> h$_0$.[r] == h$_1$.[r])
\end{lstlisting}

\iffull
Our first equivalence, listed below, shows that if a command's read and write
footprints are disjoint, then it is idempotent.
The proofs of \ls$idem$ and the other lemmas below are perhaps
peculiar to SMT-based proofs. In all cases, the proofs involve simply
mentioning the terms \ls$reify (c ()) h$, which suffice to direct the
SMT solver's quantifier instantiation engine towards finding a
proof. While more explicit proofs are certainly possible, with
experience, concise SMT-based proofs can be easier to write.

\begin{lstlisting}
let idem #rs #ws (c:cmd rs ws): 
  Lemma (requires (disjoint rs ws)) (ensures ((c $\bindseq$ c) $\sim$ c))
  = forall_intro (fun h -> let (), h$_1$ = reify (c ()) h in 
      let _ = reify (c ()) h$_1$ in () 
      <: Lemma (equiv_on_h (c $\bindseq$ c) c h))
\end{lstlisting}

Our next equivalence shows that two commands can be swapped if they
\else
For instance, we can prove that two commands can be swapped if they
\fi
write to disjoint sets, and if the read footprint of one does not
overlap with the write footprint of the other---this lemma is
identical to a rule for swapping commands in a logic presented
by \citet{BartheGB09}.

\begin{lstlisting}
let swap #rs1 #rs2 #ws1 #ws2 (c1:cmd rs1 ws1) (c2:cmd rs2 ws2)
  :Lemma (requires (disjoint ws1 ws2 /\ disjoint rs1 ws2 /\ 
                   disjoint rs2 ws1))
          (ensures ((c1 $\bindseq$ c2) $\sim$ (c2 $\bindseq$ c1)))
  = forall_intro (fun h -> let _ = reify (c1 ()) h, reify (c2 ()) h in 
    () <: Lemma (equiv_on_h (c1 $\bindseq$ c2) (c2 $\bindseq$ c1) h))
\end{lstlisting}

\iffull

Next, we show elimination of redundant writes by proving that \ls{c1 $\bindseq$ c2}
is equivalent to \ls{c2} if \ls{c1}'s write footprint is \iffull (a) \fi a
subset of \ls{c2}'s write footprint, and \iffull (b) \fi disjoint from \ls{c2}'s
read\iffull footprint\else{s}\fi.

\begin{lstlisting}
let redundant$\_$writes #rs1 #rs2 #ws1 #ws2 
  (c1:cmd rs1 ws1) (c2:cmd rs2 ws2)
  : Lemma (requires (disjoint ws1 rs2 $\wedge$ ws1 $\subseteq$ ws2))
          (ensures  ((c1 $\bindseq$ c2) $\sim$ c2))
  = forall_intro (fun h -> let _ = reify (c1 ()) h, reify (c2 ()) h in 
      () <: Lemma (equiv_on_h (c1 $\bindseq$ c2) c2 h))
\end{lstlisting}

\else
The extended version \cite{relational} also verifies command idempotence and
elimination of redundant writes.

\fi




\subsection{Relational Hoare Logic}
\label{sec:benton2004-rhl}

Beyond generic footprint-based transformations, one may also prove
program-specific equivalences. Several logics have been devised for
this, including,
e.g., \citepos{benton04relational} Relational Hoare logic (RHL). We
show how to derive RHL within our framework by proving the soundness
of each of its rules as lemmas about a program's reification.

\paragraph{Model}
To support potentially diverging computations, we instrument
shallowly-embedded effectful
computations with a \emph{fuel} argument, where the value of the fuel
is irrelevant for the behavior of a terminating computation.
\ch{So what does fuel measure? Is it recursive calls or something
  else? No trivial notion of execution steps with a shallow embedding.}
\begin{lstlisting}
type comp = f: (fuel:nat -> St bool)
   { forall h fuel fuel' . fst (reify (f fuel) h) == true /\ fuel' > fuel
     ==> reify (f fuel') h == reify (f fuel) h }
let terminates_on c h = exists fuel . fst (reify (c fuel) h) == true
\end{lstlisting}
We model effectful expressions whose evaluation always terminates and
does not change the memory state, and assignments, conditionals,
sequences of computations, and potentially diverging while
loops.\ch{``... by induction on the fuel'' This phrase makes no sense
  to me and confused POPL reviewer}

\paragraph{Deriving RHL}
An RHL judgement `\ls@related c$_1$ c$_2$ pre post@'
(where \ls@c$_1$, c$_2$@ are effectful computations, and
\ls$pre, post$ are relations over memory
states) means that the executions of \ls@c$_1$, c$_2$@ starting
\iffull(respectively)\fi in memories \ls@h$_1$, h$_2$@ related by
\ls$pre$, both diverge or both terminate with memories \ls@h$_1$', h$_2$'@
related by \ls$post$.

\begin{lstlisting}
let related (c1 c2 : comp) (pre post: (heap -> heap -> prop)) =
 (* if precondition holds on initial memory states, then *)
 forall h1 h2 . pre h1 h2 ==> 
 (* c1 and c2 both terminate or both diverge, and *)
 ((c1 `terminates_on` h1 <==> c2 `terminates_on` h2) /\ 
  (forall fuel h1' h2' . (reify (c1 fuel) h1 == (true, h1') /\ 
    reify (c2 fuel) h2 == (true, h2')) ==> (* if both terminate, *)
   post h1' h2')) (* postcondition holds on final memory states *)
\end{lstlisting}

From these reification-based definitions, we prove every rule of RHL. Of
the 20 rules and equations of RHL presented
by \citet{benton04relational}, 16 need at most 5 lines of proof
annotation each, among which 10 need none and are proven
automatically. Rules related to while loops often require some manual
induction on the fuel.
\iffull
Thus, modeling computations, program
logic rules, and their soundness proofs amount about 1500 lines of F*
code overall.\ch{Do we really need this given the table?}
\fi

\iffull
\begin{lstlisting}
(* Example of fully automatic soundness proof: dead while *)
let r_dwhll ($b$: exp bool) ($c$: computation) 
  ($\Phi$: (heap -> heap -> prop)) : Lemma
    (ensures (related (while $b$ $c$) skip $(\Phi \land \lnot b_{\text{left}})$ $(\Phi \land \lnot b_{\text{left}})$)) = ()
\end{lstlisting}
\fi

With RHL in hand, we can prove program equivalences
applying syntax-directed rules, focusing the intellectual effort
on finding and proving inductive invariants to relate loop bodies.
When RHL is not powerful enough, we can escape back to
the reification of commands to complete a direct proof
in terms of the operational semantics.
\ifcamera
In the extended version \cite{relational} we sketch a
program-specific equivalence built using our embedding of RHL
in \fstar.
\fi

\iffull
\paragraph{Example}
Following \citet{benton04relational}, we prove an example hoisting an
assignment out of a loop:
\[
\newcommand\defPhi{\begin{array}{l}I_{\text{left}} = I_{\text{right}} \land \\ N_{\text{left}} = N_{\text{right}} \land \\ Y_{\text{left}} = Y_{\text{right}} \end{array}}
\newcommand\defas[2]{\begin{tabular}{l} \fcolorbox{red}{white}{#1} \\ {\textcolor{red}{#2}} \end{tabular}}
\begin{array}{rcl}
\vdash \defas{\begin{tabular}{l} $\mathsf{while} ~ (I < N)$ \\ \hspace{1em} $X := Y + 1;$ \\ \hspace{1em} $I := I + X$ \end{tabular}}{$L$} &\rightsquigarrow& \defas{\begin{tabular}{l} $X := Y + 1;$ \\ $\mathsf{while} ~ (I < N)$ \\ \hspace{1em} $I := I + X$ \end{tabular}}{$R$} : \\ \defas{$\defPhi$}{$\Phi$} &\Rrightarrow& \defPhi

\end{array}
\]
In other words, the judgement above preserves the invariant $\Phi$ stating that the two programs $L$ and $R$  compute the same values for $I, N, Y$, with $X$ being neglected (which is already useful enough if $X$ is known to be dead in the code following the while loops).

\begin{lstlisting}
let proof () : Lemma (ensures (related $L$ $R$ $\Phi$ $\Phi$)) =
  (* intermediate invariants for the loop bodies *)
  let $\Phi_1$ = $\Phi \land (X_{\text{right}} = Y_{\text{right}} + 1)$ in 
  let $\Phi_2$ = $\Phi_1 \land (X_{\text{left}} = X_{\text{right}})$ in
  assert (related skip (assign $X$ $(Y + 1)$) $\Phi$ $\Phi_1$); (* dead assign *)
  assert (related (assign $X$ $(Y + 1)$) skip $\Phi_1$ $\Phi_2$); (* dead assign *)
  assert (related (assign $I$ $(I + X)$) (assign $I$ $(I + X)$) $\Phi_2$ $\Phi_2$); (* assign *)
  assert (related (seq (assign $X$ $(Y + 1)$) (assign $I$ $(I + X)$)) 
                  (assign i $(I + X)$) $\Phi_1$ $\Phi_2$); (* seq, elim. skip *)
  r_while $(I < N)$ $(I < N)$ (seq (assign $X$ $(Y + 1)$) (assign $I$ $(I + X)$)) 
          (assign $I$ $(I + X)$) $\Phi_1$;
  (* seq, elim. skip *)
  assert (related $L$ (while $(I < N)$ (assign $I$ $(Y + 1)$)) $\Phi_1$ $\Phi$) 
\end{lstlisting}
\[
  \inferrule[ $\mathsf{r\_while} ~ B ~ B' ~ C ~ C' ~ \Phi : $]{
  \vdash C \rightsquigarrow C' : \Phi \land B_{\text{left}} \land B'_{\text{right}} \Rrightarrow \Phi \land (B_{\text{left}} = B'_{\text{right}})
}
{
  \vdash \mathsf{while} ~ B ~ \mathsf{do} ~ C \rightsquigarrow \mathsf{while} ~ B' ~ \mathsf{do} ~ C' : \Phi \land (B_{\text{left}} = B'_{\text{right}}) \Rrightarrow \\  \Phi \land \lnot (B_{\text{left}} \lor B'_{\text{right}})
}
\]
The proof shows that applications of RHL rules (including dead assignment rules) are
actually syntax-directed, so that the only nontrivial effort needed is
to provide the intermediate verification condition relating the bodies
of the loops.

In more detail, for a given proposition $\phi$, \ls!assert $\phi$!
tries to prove $\phi$ and, if successful, adds $\phi$ to the proof
context as a fact that can be automatically reused by the later parts
of the proof. To prove $\phi$, proof search relies not only on the
current proof context,
but also on those lemmas in the global context
that are associated with \emph{triggering patterns}: if the shape of
$\phi$ matches the triggering pattern of some lemma $f$ in the global
context, then $f$ is applied (\emph{triggered}) and the proof search
recursively goes on with the preconditions of $f$. This proof search
is actually performed by the Z3 SMT solver
through \emph{e-matching} \cite{MouraB07}.

In our example proof, 
\ls!assert (related skip (assign $X$ $(Y + 1)$) $\Phi$ $\Phi_1$)!
tries to prove that an assignment can be erased; based on the syntax
of both commands of the relation, e-matching successfully selects the
corresponding dead assignment rule of RHL. In fact, this \ls!assert!
also allows specifying the intermediate condition $\Phi_1$ that is to
be used to verify the rest of the bodies of $L$ and $R$, which cannot
always be guessed by proof search. Alternatively, the user can also
explicitly apply an RHL rule by directly calling the corresponding
lemma, which is illustrated by the call to \ls!r_while! to prove that
the two while loops are related. In that case, the postcondition of
the lemma is added to the proof context for the remainder of the
proof. This way, the user can avoid explicitly spelling out the fact
proven by the lemma; moreover, since the lemma to apply is explicitly
given, the SMT solver only has to prove the preconditions of
the lemma, if any.

This example is 33 lines of F* code and takes 25 seconds to
check. This time could be improved substantially. However, perhaps
more interesting, this experiment suggests developing tactics to
automatically use Benton's RHL whenever possible, while still keeping
the possibility to escape back to semantic approaches wherever RHL is
not powerful enough. We leave this as future work.
\fi

\section{Cryptographic security proofs}
\label{sec:crypto}

We show how to construct a simple model for reasoning about
probabilistic programs that sample values from discrete
distributions. In this model, we prove the soundness of rules of
probabilistic Relational Hoare Logic (pRHL)~\citep{BartheGB09}
allowing one to derive (in-)equalities on probability quantities from
pRHL judgments. We illustrate our approach by formalizing \iffull two \else a \fi  simple
cryptographic proof\iffull{s}\fi: the perfect secrecy of one-time pad encryption
\iffull and a crucial lemma used by \citet{BartheGB09}
in the proof of semantic security of ElGamal encryption \fi .

The simplicity of our examples pales in comparison with complex proofs
formalized in specialized tools based on pRHL like
EasyCrypt~\citep{BartheGB12} or FCF~\citep{PetcherM15}, yet our
examples hint at a way to prototype and explore proofs in pRHL with a
low entry cost.

\subsection{A monad for random sampling}

We begin by defining a monad for sampling from the uniform
distribution over bitvectors of a fixed length \ls$q$. We implement
the monad as the composition of the state and exception monads where the
state is a finite tape of bitvector values together with a pointer to
a position in the tape.
The \ls$RAND$ effect provides a single action, \ls$sample$, which
reads from the tape the value at the current position and advances the
pointer to the next position, or raises an exception if the pointer is
past the end of the tape.
\begin{lstlisting}
type value = bv q
type tape = seq value
type id = i:$\nat${i < size}
type store = id * tape
type rand a = store -> M (option a * id)
total new_effect {
  RAND: a:Type -> Effect
  with repr = rand a;
       bind = fun (a b:Type) (c:rand a) (f:a -> rand b) s ->
                let r, next = c s in
                match r with
                | None   -> None, next
                | Some x -> f x (next, snd s);
       return = fun (a:Type) (x:a) (next,_) -> (Some x, next);
       sample = fun () s -> let next, t = s in
                  if next + 1 < size then (Some (t n), n + 1) 
                  else (None, n) }
effect Rand a = RAND a (fun initial_tape post -> forall x. post x)
\end{lstlisting}

Assuming a uniform distribution over initial tapes, we define the
unnormalized measure of a function \ls{$p$:a -> $\nat$} with respect
to the denotation of a reified computation in \ls{$f$:Rand a} as
\ls$let mass f p = sum (fun t -> let r,_ = f (0, t) in p r)$
where \ls'sum: (tape -> $\nat$) -> $\nat$' is the summation operator over
finite tapes.
When $p$ only takes values in $\{0,1\}$, it can be regarded as an
\emph{event} whose probability with respect to the distribution generated by
$f$ is
$$
\Pr[ f : p ]
  = \frac{1}{|\mathsf{tape}|} \times \sum_{t\ \in\ \mathsf{tape}} p\ (\mathsf{fst}\ (f\ t))
  = \frac{\mathsf{mass}\ f\ p}{|\mathsf{tape}|}
$$
We use the shorthand
$\Pr[f = v] = {|\mathsf{tape}|}^{-1} \times \mathsf{mass}\ f\ (\mathsf{point}\ v)$
for the probability of a successful computation returning a value $v$,
where
\ls{let point x = fun y -> if y = Some x then 1 else 0}.

\subsection{Perfect secrecy of one-time pad encryption}

The following effectful program uses a one-time key
\ls$k$ sampled uniformly at random to encrypt a bitvector $m$:
\begin{lstlisting}
let otp (m:value) : Rand value = let k = sample () in m $\oplus$ k
\end{lstlisting}
We show that this construction, known as \emph{one-time pad}, provides
\emph{perfect secrecy}. That is, a ciphertext does not give away any
information about the encrypted plaintext, provided the encryption key
is used just once. Or equivalently, the distribution of the one-time
pad encryption of a message is independent of the message itself,
$\forall m_0,~m_1,~c.\ \Pr[\mathsf{otp}\ m_0 = c] = \Pr[\mathsf{otp}\ m_1 = c]$.
We prove this by applying two rules of pRHL, namely [R-Rand] and
[PrLe].\ch{Important question: have we actually proven these rules?}
The former allows us to relate the results of two
probabilistic programs by showing a bijection over initial random
tapes that would make the relation hold (intuitively, permuting
equally probable initial tapes does not change the resulting
distribution over final tapes). The latter allows us to infer a
probability inequality from a proven relation between probabilistic
programs. Together, the two rules allow us to prove the following lemma:
\begin{lstlisting}
val mass_leq: #a:Type -> #b:Type -> 
  c1:(store -> M (a * id)) -> c2:(store -> M (b * id)) ->
  p1:(a -> nat) -> p2:(b -> nat) -> bij:bijection -> Lemma
  (requires (forall t. let r1,_ = c1 (to_id 0,t) in 
               let r2,_ = c2 (to_id 0,bij.f t) in p1 r1 <= p2 r2))
  (ensures (mass c1 p1 <= mass c2 p2))
\end{lstlisting}
The proof is elementary from rearranging terms in summations according
to the given bijection.
The following secrecy proof of one-time pad is immediate from this
lemma using as bijection on initial tapes \iffull the function \fi
\ls{fun t -> upd t 0 (t 0 $\oplus$ m0 $\oplus$ m1)}:
\begin{lstlisting}
val otp_secure: m0:value -> m1:value -> c:value -> Lemma
  (let f0, f1 = reify (otp m0), reify (otp m1) in 
   mass f0 (point c) == mass f1 (point c))
\end{lstlisting}

\iffull
\subsection{A step in the proof of semantic security of ElGamal encryption}

Another example following a similar principle is a probabilistic
equivalence used in the proof of semantic security of ElGamal
encryption by \citepos{BartheGB09}. This equivalence, named
\texttt{mult\_pad} in that paper, proves the independence of the
adversary's view from the hidden bit $b$ that the adversary has to
guess in the semantic security indistinguishability game, and thus
shows that the adversary cannot do better than a random guess.

ElGamal encryption is parametric on a cyclic group of order $q$, and a
generator $g$.
%
%
Roughly stated, the equivalence says that if one applies the group
operation to a uniformly distributed element of the group and some
other element, the result is uniformly distributed, that is
$\Rand{z}{\zq}; \zeta \leftarrow g^z \times m_b$
and
$\Rand{z}{\zq}; \zeta \leftarrow g^z$
induce the same distribution on $\zeta$ (which is thus independent of
$b$).
To prove this, we modify the \ls{RAND} effect
to use random tapes of elements of $\zq$ rather than bitvectors, an define
\begin{lstlisting}
let elgamal$_0$ (m:group) : Rand group = let z = sample () in g^z
let elgamal$_1$ (m:group) : Rand group = let z = sample () in (g^z) * m
\end{lstlisting}
and prove, again using \ls{mass_leq}, the following lemma
\begin{lstlisting}
val elgamal_equiv: m:group -> c:group -> Lemma
  (let f1, f2 = reify (elgamal$_0$ m), reify (elgamal$_1$ m) in 
  mass f1 (point c) == mass f2 (point c))
\end{lstlisting}

\fi

\section{Information-flow control}
\label{sec:ifc}




In this section, we present a case study examining various styles of
information-flow control (IFC), a security paradigm based on
\emph{noninterference} \cite{goguen82security}, a property that
compares two runs of a program differing only in the program's secret
inputs and requires the non-secret outputs to be equal. Many
special-purpose systems, including syntax-directed type systems, have
been devised to enforce noninterference-like security properties
\cite[see, e.g.,][]{SabelfeldMyers06IFC, HedinS12}.

We start our IFC case study by encoding a classic IFC type system 
\citep{volpano1996ifc} for a small deeply-embedded imperative
language and proving its correctness (\autoref{sec:ifc-while}).
In order to augment the permissiveness of our analysis we then show how to
compose our IFC type system with precise semantic proofs
(\autoref{sec:ifc-combining}).
\iffull
As IFC is often too strong for practical use, the final step in our IFC case
study is a semantic treatment of declassification based on delimited release
\citep{SabelfeldMyers03DL} (\autoref{sec:declassification}). An additional case
study on a runtime monitor for IFC is presented in \autoref{sec:ifc-monitor}.
\else
In the extended version~\cite{relational} we additionally treat a runtime monitor or IFC and 
delimited release. 
\fi
We conclude that our method for relational
verification is flexible enough to accommodate various IFC
disciplines, allowing comparisons and compositions within the same
framework.

\subsection{Deriving an IFC type system}
\label{sec:ifc-while}

\iffull
\begin{figure}
\def\MathparLineskip{\lineskip=0.1cm}
  \begin{mathpar}
\inferrule[ESub]{
  \Gamma |- e : l_1 \\ l_1 \leq l_2
}
{
  \Gamma |- e : l_2
}
\and
\inferrule[EVar]{}
{
  \Gamma |- r : \Gamma \left( r \right)
}
\and
  \inferrule[EInt]{i:\text{int}
}
{
  \Gamma |- i : \text{L}
}
\and
  \inferrule[EBinOp]{\Gamma |- e_1 : l \\ \Gamma |- e_2 : l
}
{
  \Gamma |- e_1 \oplus  e_2 : l
}
\and
  \inferrule[CSub]{\Gamma,\text{pc}:l_1 |- c \\ l_2 \leq l_1
}
{
  \Gamma, \text{pc}:l_2 |- c
}
\and
  \inferrule[CAssign]{\Gamma |- e : \Gamma \left( r \right)
}
{
  \Gamma, \text{pc}:\Gamma \left( r \right) |- r := e
}
\and
  \inferrule[CSeq]{\Gamma,\text{pc}:l |- c_1 \\ \Gamma,\text{pc}:l |- c_2
}
{
  \Gamma, \text{pc}:l |- c_1; c_2
}
\and
  \inferrule*[left=CCond]{\Gamma |- e : l \\ \Gamma,\text{pc}:l |- c_1 \\ \Gamma,\text{pc}:l |- c_2
}
{
  \Gamma, \text{pc}:l |- \text{if } e = 0 \text{ then } c_1 \text{ else } c_2
}
\and
  \inferrule*[left=CWhile]{\Gamma |- e : l \\ \Gamma,\text{pc}:l |- c  \\ \Gamma |- e' : l'
}
{
  \Gamma, \text{pc}:l |- \text{while } e \neq 0 \text{ do } c \left(\text{decr } e'\right)
}
\and
  \inferrule*[left=CSkip]{}
{
  \Gamma, \text{pc}:\text{H} |- \text{skip}
}
\end{mathpar}
\caption{A classic IFC type system}
\label{fig:ts_ifc}
\end{figure}

\else

\begin{figure}
\def\MathparLineskip{\lineskip=0.1cm}
  \begin{mathpar}
  \inferrule[CSub]{\Gamma,\text{pc}:l_1 |- c \\ l_2 \leq l_1
}
{
  \Gamma, \text{pc}:l_2 |- c
}
\and
  \inferrule[CAssign]{\Gamma |- e : \Gamma \left( r \right)
}
{
  \Gamma, \text{pc}:\Gamma \left( r \right) |- r := e
}
\and
  \inferrule*[left=CCond]{\Gamma |- e : l \\ \Gamma,\text{pc}:l |- c_1 \\ \Gamma,\text{pc}:l |- c_2
}
{
  \Gamma, \text{pc}:l |- \text{if } e = 0 \text{ then } c_1 \text{ else } c_2
}
\end{mathpar}
\caption{A classic IFC type system (selected rules)}
\label{fig:ts_ifc}
\end{figure}
\fi

Consider the following small \emph{while} language consisting of
expressions, which may only read from the heap, but not modify it, and
commands,  which may write to the heap and branch, depending on its contents.
The definition of the language
should be unsurprising, the only subtlety worth noting is
the \ls$decr$ expression in the while command, a metric used to ensure
loop termination. \ch{Why? If this is explained latter need at
  least a forward reference. I don't think it is though; it's a very
  mysterious and non-standard thing though that needs serious
  justification. And it is caused by a limitation in our F*-based
  method that we should at some point explain: we can't reason
  extrinsically about non-terminating programs.}
\ch{I wonder whether Tahina's fuel technique or Kenji's partiality
  monad would be better ways to approach this.}
\ch{Alternatively, for loops instead of while loops?}

\[
  \begin{array}{r c l}
    e & ::= &  i ~|~ r ~|~ e_1 \oplus e_2 \\
    c & ::= &  \text{skip} ~|~ r := e ~|~ c_1 ;c_2 ~|~ \text{if } e = 0 \text{ then } c_1 \text{ else } c_2 \\
    & & ~|~ \text{while } e \neq 0 \text{ do } c \left(\text{decr } e'\right)
  \end{array}
\]
\paragraph*{A classic IFC type system}
\citet{volpano1996ifc} devise an IFC type system \iffull for a similar
language \fi to check that programs executing over a memory containing
both secrets (stored in memory locations labeled \ls$High$) and
non-secrets (in locations labeled \ls$Low$) never leak secrets into
non-secret locations. The type system includes two judgments $\Gamma
|- e : l$, which states that the expression \ls$e$ (with free
variables in $\Gamma$) depends only on locations labeled $l$ or
lower; and $\Gamma, \text{pc}:l |- c$, which states that a command $c$
in a context that is \emph{control-dependent} on the contents of
memory locations labeled $l$, does not leak secrets. \iffull The main 
\else Some selected rules \fi of
their system, as adapted to our example language, are shown in
Figure~\ref{fig:ts_ifc}.

\iffull
Our goal in this section is to embed this \emph{while} language
in \fstar, to define an interpreter for it, and to derive Volpano et
al.'s type system by relating multiple runs of the interpreter.
In doing so, we highlight several distinctive features of our
approach, including the use of multiple monads to structure our
interpreter and simplify our proofs.
\ch{Probably don't have the space for such outlines}
\fi

\paragraph*{Multiple effects to structure the \emph{while} interpreter}
We deeply embed the syntax of \emph{while} in \fstar using
data types \ls$exp$ and \ls$com$, for expressions and commands,
respectively.
The expression interpreter \ls$interp_exp$ only requires reading the
value of the variables from the \ls$store$, whereas the command
interpreter, \ls$interp_com$, also requires writes to the
\ls$store$, where \ls$store$ is an integer store mapping
a fixed set of integer references `\ls$ref int$' to
\ls$int$.
Additionally, \ls$interp_com$ may also raise an \ls$Out_of_fuel$
exception when it detects that a loop may not terminate (e.g., because
the claimed metric is not actually decreasing).\ch{Is this an
  out of fuel exception (as the current name suggests) or a
  wrong metric exception (as the explanation suggests)?}
We could define both interpreters using a single effect, but this
would require us to prove that \ls$interp_exp$ does not change the
store and does not raise exceptions.
Avoiding the needless proof overhead, we use a \ls$Reader$ monad
for \ls$interp_exp$ and \ls$StExn$, a combined state and exceptions
monad, for \ls$interp_com$. By defining \ls$Reader$ as a \ls$sub_effect$
of \ls$StExn$, expression interpretation is transparently lifted
by \fstar{} to the larger effect when interpreting commands.
\iffull
\begin{lstlisting}
type reader (a:Type) = store -> Tot a
total new_effect { READER : a:Type -> Effect
  with repr = reader;
    return = fun (a:Type) (x:a) (s:store) -> x;
    bind = fun (a b : Type) (f:reader a) (g: a -> reader b) (s:store) -> 
                let z = f s in g z s; get = fun () (s:store) -> s }
type stexn (a:Type) = store -> Tot (either a exn * store)
total new_effect { STEXN $\ldots$ }
sub_effect READER ~> STEXN 
  { lift = fun (a:Type) (f:reader a) (s:store) -> let x = f s in (Inl x, s)  }
\end{lstlisting}
\fi
Using these effects, \ls$interp_exp$ and \ls$interp_com$ form a
standard, recursive, definitional interpreter for \emph{while}, with
the following trivial signatures. \iffull Just as we sometimes use \ls$St$,
the unindexed version of \ls$STATE$, here we use \ls$Reader$
and \ls$StExn$, unindexed versions of \ls$READER$ and \ls$STEXN$ with
simple pre- and postconditions.
\fi

\begin{lstlisting}
val interp_exp: exp -> Reader int
val interp_com: com -> StExn unit
\end{lstlisting}


\iffull
Similarly, the memoization example from \autoref{sec:memo} uses an
effect that is specialized to the target application: a state monad
where the state is a partial finite map storing all arguments on which
a particular function was called and what answers it returned.

\fi


%
%


\paragraph*{Deriving IFC typing for expressions}
For starters, we use a \ls$store_labeling = ref int -> label$,
where \ls@label@ $\in$ \ls@{High, Low}@, to partition the store between
secrets (\ls$High$) and non-secrets (\ls$Low$).
An expression is noninterferent at level $l$ when its interpretation
does not depend on locations labeled greater than $l$ in the store.
To formalize this, we define a notion of \emph{low-equivalence} on
stores, relating stores that agree on the contents of
all \ls$Low$-labeled references, and noninterferent expressions (at
level \ls$Low$, i.e., \ls$ni_exp env e Low$) as those whose
interpretation is identical in low-equivalent stores.

\begin{lstlisting}
type low_equiv (env:store_labeling) (s0 s1:store) =
    forall (r:ref int). env r=Low ==> s0.[r] == s1.[r]
let ni_exp (env:store_labeling) (e:exp) (l:label) =
    forall (s0 s1:store). (low_equiv env s0 s1 /\ l == Low) ==> 
      reify (interp_exp e) s0 == reify (interp_exp e) s1
\end{lstlisting}





With this definition of noninterference for expressions we capture
the semantic interpretation of the typing judgment
$\Gamma |- e : l$: if the expression $e$ can be assigned the label \ls$Low$, 
then the computation of $e$ is only influenced by \ls$Low$ values.
\iffull
Using this definition, we can derive the expression rules of
Figure~\ref{fig:ts_ifc}; for instance here is a lemma for the \textsc{EBinOp} rule:

\begin{lstlisting}
let binop_exp (env:store_labeling) (op:binop) (e1 e2:exp) (l:label)
 : Lemma (requires (ni_exp env e1 l /\ ni_exp env e2 l)) 
         (ensures (ni_exp env (AOp op e1 e2) l)) = ()
\end{lstlisting}

We construct a lemma from the inference rule in a straightforward
manner: the premise of the inference rule forms the \ls$requires$
clause, while the conclusion of the rule forms the \ls$ensures$ clause.
The proof for this lemma is simple and can be discharged
purely by SMT, without the need of any further annotations. The other
rules for expressions can be shown in the same way and all of them can
be discharged by SMT.
\fi

\paragraph*{Deriving IFC typing for commands} As explained previously, the
judgment $\Gamma, \text{pc}:l |- c$ deems $c$ noninterferent when run
in context control-dependent only on locations whose label is at most
$l$. More explicitly, the judgment establishes the following two
properties:
(1) locations labeled below $l$ are not modified by $c$---this is
  captured by \ls$no_write_down$, a unary property;
(2) the command $c$ does not leak the contents of a \ls$High$ location to \ls$Low$
  location---this is captured by \ls$ni_com'$, a binary property.


\begin{lstlisting}
let run c s = match reify (interp_com c) s with
  | Inr Out_of_fuel, _ -> Loops  | _, s' -> Returns s'
let no_write_down env c l s = match run c s with
  | Loops -> True    | Returns s' -> forall (i:id). env i < l ==> s'.[i] == s.[i]
let ni_com' env c l s0 s1 = match run c s0, run c s1 with
  | Returns s0', Returns s1' -> low_equiv env s0 s1 ==> 
      low_equiv env s0' s1'
  | Loops, _ | _, Loops -> True
\end{lstlisting}

The type system is termination-insensitive,
meaning that a program may diverge depending on the value of a
secret. Consider, for instance, two runs of the program
\texttt{while hi <> 0 do \{skip\}; lo := 0}, one with \texttt{hi = 0}
and another with \texttt{hi = 1}. The first run terminates and writes
to \texttt{lo}; the second run loops forever. As such, we do not
expect to prove noninterference in case the program loops.
\iffull

\fi
Putting the pieces together, we define
$\Gamma, \text{pc}:l |- c$ to be \ls@ni_com $\Gamma$ $c$ $l$@.
\begin{lstlisting}
let ni_com (env:store_labeling) (c:com) (l:label) = 
  (forall s0 s1.  ni_com' env c l s0 s1) /\ (forall s.  no_write_down env c l s)
\end{lstlisting}

As in the case of expression typing, we derive each rule of the
command-typing judgment as a lemma about \ls$ni_com$. For example,
here is the statement for the \textsc{CCond} rule:

\begin{lstlisting}
val cond_com (env:store_labeling)(e:exp)(ct:com)(cf:com)(l:label)
: Lemma (requires (ni_exp env e l /\ ni_com env ct l 
                               /\ ni_com env cf l))
        (ensures  (ni_com env (If e ct cf) l))
\end{lstlisting}

\noindent The proofs of many of these rules
are partially automated by SMT---they take about 250 lines of
specification and proof in \fstar. Once proven, we use these rules to
build a certified, syntax-directed typechecker for \emph{while}
programs that repeatedly applies these lemmas to prove that
a program satisfies \ls$ni_com$. This \iffull certified \fi typechecker has the
following type:
\begin{lstlisting}
val tc_com : env:store_labeling -> c:com -> 
  Exn label (requires True) (ensures fun Inl l -> ni_com env c l | _ -> True)
\end{lstlisting}



\subsection{Combining syntactic IFC analysis with semantic noninterference proofs}
\label{sec:ifc-combining}

Building on \autoref{sec:ifc-while}, we show how programs that fall
outside the syntactic information-flow typing discipline can be proven
secure using a combination of typechecking and
semantic proofs of noninterference. This example is evocative (though
at a smaller scale) of the work of \citet{KustersTBBKM15}, who combine
automated information-flow analysis in the Joana analyzer
\cite{HammerS09} with semantic proofs
in the KeY verifier for Java programs \cite{DarvasHS05, SchebenS11}.
In contrast, we sketch a combination of syntactic and semantic
proofs of relational properties in {\em a single} framework.
Consider the following \emph{while} program, where the label of \ls$c$
and \ls$lo$ is Low and the label of \ls$hi$ is High.
\begin{lstlisting}[language=caml]
  while c <> 0 do hi := lo + 1; lo := hi + 1; c := c $-$ 1 (decr c)
\end{lstlisting}
The assignment \ls$lo := hi + 1$ is ill-typed in the type system
of \S\ref{sec:ifc-while}, since it directly assigns a \ls$High$
expression to a \ls$Low$ location.
However, the previous command overwrites \ls$hi$ so that
\ls$hi$ does not contain a \ls$High$ value anymore at that point.
As such, even though the IFC type system cannot prove it, the program
is actually noninterferent.
To prove it, one could directly attempt to prove \ls$ni_com$ for the
entire program, which would require a strong enough (relational)
invariant for the loop. A simpler approach is to prove just the
sub-program \ls$hi := lo + 1; lo := hi + 1$ (\ls$c_s$) 
noninterferent, while relying on the
type system for the rest of the program.
The sub-program can be automatically proven secure:

\begin{lstlisting}
let c_s_ni () : Lemma (ni_com env c_s Low) = ()
\end{lstlisting}

\noindent This lemma has exactly the form of the other standard,
typing rules proven previously, except it is specialized to the
command in question. As such, \ls$c_s_ni$ can just be used in place
of the standard sequence-typing rule (\textsc{CSeq}) when proving the
while loop noninterferent.

We can even modify our automatic typechecker from \autoref{sec:ifc-while} 
to take as input a list of commands that are already proved
noninterferent (by whichever means), and simply look up the command it
tries to typecheck in the list before trying to typecheck it syntactically.
The type (and omitted implementation) of this typechecker is very
similar to that of \ls$tc_com$, the only difference is the extra list argument:
\begin{lstlisting}
val tc_com_hybrid : env:store_labeling -> c:com -> 
  list (cl:(com*label){ni_com env (fst cl) (snd cl)}) ->
  Exn label (ensures fun ol -> Inl? ol ==> ni_com env c (Inl?.v ol))
\end{lstlisting}
We can complete the noninterference proof automatically by passing the
\ls$(c_s, Low)$ pair proved in \ls$ni_com$ by lemma
\ls$c_s_ni$ (or directly by SMT) to this hybrid IFC typechecker:
\begin{lstlisting}
let c_loop_ni () : Lemma (ensures ni_com env c_loop Low) =
 c_s_ni(); ignore (reify (tc_com_hybrid env c_loop [c_s, Low]) ())
\end{lstlisting}
Checking this in \fstar{} works by simply evaluating the invocation
of \ls$tc_com_hybrid$; this reduces fully to \ls$Inl Low$ and
the intrinsic type of \ls$tc_com_hybrid$ ensures the postcondition.







\iffull
\subsection{Semantic declassification}
\label{sec:declassification}

Beyond noninterference, reasoning directly about relational properties
allows us to characterize various forms of \emph{declassification}
where programs intentionally reveal some information about
secrets. For example, \citet{SabelfeldMyers03DL}
propose \emph{delimited release}, a discipline in which programs are
allowed to reveal the value of only certain pure expressions.

In a simple example by Sabelfeld and Myers
some amount of money (\ls$k$) is transferred
from one account (\ls$hi$) to another (\ls$lo$). Simply by
observing whether or not the funds are received, the owner of the
\ls$lo$ account gains some information about the other account, namely
whether or not \ls$hi$ contained at least \ls$k$ units of
currency---this is, however, by design.

\begin{lstlisting}
let transfer (k:int) (hi:ref int) (lo:ref int) = 
  if k < !hi then (hi := !hi $-$ k; lo := !lo + k)
\end{lstlisting}

To characterize this kind of intentional release of information,
delimited release describes two runs of a
program in initial states where the secrets, instead of being
arbitrary, are related in some manner, e.g., the initial states
agree on the value of the term being explicitly declassified. This is
easily captured in our setting. For example, we can prove the
following lemma for \ls$transfer$, which shows that \ls$lo$ gains no
more information than intended.


\begin{lstlisting}
let transfer_ok (k:int) (hi lo:ref int{addr_of lo <> addr_of hi}) 
  (s0 s1:heap{lo $\in$ s0 /\ hi $\in$ s0 /\ lo $\in$ s1 /\ hi $\in$ s1}) : Lemma
     (* initial memories agree on lo and on the declassified term *)
     (requires (s0.[lo] == s1.[lo] /\ (k < s0.[hi] <==> k < s1.[hi]))) 
     (ensures ((snd (reify (transfer k hi lo) s0)).[lo] == 
               (snd (reify (transfer k hi lo) s1)).[lo])) = ()
\end{lstlisting}


Delimited release was about the {\em what} dimension of
declassification \cite{SabelfeldS09}.
We also built a very simple model that is targeted at the {\em when}
dimension, illustrating a customization of the monadic model to the
target relational property.
For instance, to track when information is declassified, we augment
the state with a bit recording whether the secret component of the
state was declassified and is thus allowed to be leaked.
\begin{lstlisting}
type ifc_state = { secret:int; public:int; release:bool }
new_effect STATE_IFC = STATE_h ifc_state
\end{lstlisting}
In this case the noninterference property depends on the extra
instrumentation bit we added to the state.
\begin{lstlisting}
let ni (f:unit -> St unit) =
 forall s0 s1. let (_, s0'), (_, s1') = reify (f ()) s0, reify (f ()) s1 in
  s0'.release \/ s1'.release \/ (low_equiv s0 s1 ==> low_equiv s0' s1')
\end{lstlisting}

\subsection{Soundness of an IFC monitor}
\label{sec:ifc-monitor}

\begin{figure}
  \begin{mathpar}
\inferrule[EVar]{}
{
  S, \Gamma |- r \rightarrow  \left< S \left( r \right), \Gamma \left( r \right) \right>
}
\and
  \inferrule[EInt]{i:\text{int}
}
{
  S, \Gamma |- i \rightarrow \left<i, \text{L} \right>
}
\and
  \inferrule[EBinOp]{S |- e_1 \rightarrow \left<v_1,l_1\right> \\
    S, \Gamma |- e_2  \rightarrow \left< v_2,l_2 \right>
}
{
  S, \Gamma |- e_1 \oplus  e_2 \rightarrow \left< v_1 \oplus v_2, l_1 \sqcup l_2 \right>
}
\and
  \inferrule[CAssign]{S, \Gamma |- e \rightarrow \left< v_e, l_e \right> \\
    \Gamma \left( r \right) = l_r \\\\
    l_e \sqcup \text{pc} \leq l_r
}
{
  S, \Gamma, \text{pc} |- r := e \rightarrow S\left[r \mapsto v_e\right]
}
\and
  \inferrule[CCondTrue]{S, \Gamma |- e \rightarrow \left< v_e, l_e \right> \\
    v_e = 0 \\
    S, \Gamma,(\text{pc} \sqcup l_e) |- c_1 \rightarrow S_1
}
{
  S, \Gamma, \text{pc} |- \text{if } e = 0 \text{ then } c_1 \text{ else } c_2 \rightarrow S_1
}
\end{mathpar}
\caption{Semantics of the IFC monitor}
\label{fig:ifc_monitor}
\end{figure}

Another popular technique for the enforcement of IFC are
runtime monitors: the idea is to dynamically track the security labels
of  expressions and to check them at runtime in order to detect  IFC
violations, which cause the execution to halt. Here we implement an interpreter for the while language presented in
\autoref{sec:ifc-while} extended with the security monitor proposed by
\citet{SabelfeldR09}: a selection of the semantic rules is reported in \autoref{fig:ifc_monitor}.
The store $S$ maps references to integers, while the store labeling $\Gamma$
maps references to security labels, which are then used to derive labels for
expressions. Assignments are subject to the expected security checks at
run-time.

We embed the monitor in \fstar, obtaining a machine-checked
proof of soundness for it.
The interpretation functions for expressions and commands have the following
signatures:

\begin{lstlisting}
val interp_exp_monitor: store_labeling -> exp -> Reader (int * label)
val interp_com_monitor: store_labeling -> label -> com -> StExn unit
\end{lstlisting}

We prove termination-insensitive non-interference for interpretation with the
monitor and capture this with the following lemma:

\begin{lstlisting}
val dyn_ifc (s0:store) (s1:store) (env:store_labeling) (c:com) (pc:label) :
    Lemma (requires (low_equiv env s0 s1))
      (ensures (match (reify (interp_com_monitor env pc c)) s0, 
                     (reify (interp_com_monitor env pc c)) s1 with
               | (Inl _, s0'), (Inl _, s1')-> low_equiv env s0' s1'
               | _ -> True))
\end{lstlisting}

Intuitively, we show that for any two low-equivalent initial stores,
the two resulting
stores are also low equivalent, if the interpretation with the monitor
terminates without a runtime exception.

While the result looks similar to the one shown for the type system,
there is a subtle difference in the enforced security property. Consider the
following example where the label of \ls$hi$ is High and the label of \ls$lo$
is Low:
\begin{lstlisting}[language=caml]
if (hi=0) skip else lo := 0
\end{lstlisting}
The assignment to a low reference on the else branch is leaking
information about the value of the high reference in the conditional expression.
Nevertheless, if the then-branch of the
conditional is taken, the monitor will not report a violation, as it does not
inspect the else-branch.
This example does however not break our theorem, since our theorem only relates
pairs of programs that terminate normally,
while for all stores in which the else branch
is taken, the execution of the interpreter halts with an error.
The monitor is collapsing the implicit-flow channel into an
erroneous termination channel, thereby enforcing error-insensitive
non-interference.
%
%
For comparison, notice that the (termination-insensitive)
type system from \autoref{sec:ifc-while}  accepts a
variant of the program above, in which the low assignment is replaced by a
non-terminating loop.



\fi

\section{Program optimizations and refinement}
\label{sec:equiv}
\label{sec:refinement}

This section presents two complete examples to prove a few,
classic algorithmic optimizations correct. These properties are very
specific to their application domains and a special-purpose
relational logic would probably not be
suitable. Instead, we make use of the generality of our approach to
prove application-specific relational properties (including $4$- and
$6$-ary relations) of higher-order programs with local state.  In
contrast, most prior relational logics are specialized to proving
binary relations, or, at best, properties of $n$ runs of a single
first-order program \citep{SousaD16}.

\subsection{Effect for memoizing recursive functions}
\label{sec:memo}

First, we look at memoizing total functions,
including memoizing a function's recursive calls 
based on a partiality representation
technique due to \citet{McBride15}.
We prove that a memoized function is extensionally equal to the original.

We define a custom effect \ls$Memo$, a monad with a state consisting of a
(partial, finite) mapping from a function's domain type (\ls$dom$)
to its codomain type (\ls$codom$),
with two actions:
\ls$get : dom -> Memo (option codom)$, which returns a memoized value if it exists; and
\ls$put : dom -> codom -> Memo unit$, which adds a new memoization pair
  to the state.\footnote{
This abstract model could be implemented efficiently,
for instance by an imperative hash-table with a specific
memory-management policy.}

\paragraph*{Take 1: Memoizing total functions} Our
goal is to turn a total function \ls$g$ into a memoized
function \ls$f$ computing the same values as \ls$g$.
This relation between \ls$f$'s reification and \ls$g$ is captured by
the \ls$computes$ predicate below, depending on an invariant of the
memoization state, \ls$valid_memo$.
A memoization state \ls$h$ is valid for memoizing some total
function \ls$g : (dom -> codom)$ when \ls$h$ is a subset of the graph
of \ls$g$:
\begin{lstlisting}
let valid_memo (h:memo_st) (g:dom -> codom) = 
  for_all_prop (fun (x,y) -> y == g x) h
let computes (f: dom -> Memo codom) (g:dom -> codom) =
  forall h0. valid_memo h0 g ==> (forall x. (let y, h1 = reify (f x) h0 in 
                                y == g x /\ valid_memo h1 g))
\end{lstlisting}
We have \ls$f `computes` g$ when given any state \ls$h0$ containing
a subgraph of \ls$g$, \ls$f x$ returns \ls$g x$ and maintains the
invariant that the result state \ls$h1$ is a subgraph of \ls$g$.
It is easy to program and verify a simple memoizing function:
\begin{lstlisting}
let memoize (g : dom -> codom) (x:dom) = 
  match get x with Some y -> y | None -> let y = g x in put x y; y
let memoize_computes g :Lemma ((memoize g) `computes` g) = ...
\end{lstlisting}
The proof of this lemma is straightforward: we only need to show that
the value \ls$y$ we get back from the heap in the first branch is
indeed \ls$g x$ which is enforced by the \ls$valid_memo$ in the
precondition of \ls$computes$.

\paragraph*{Take 2: Memoizing recursive calls}
Now, what if we want to memoize a recursive function, for
example, a function computing the Fibonacci sequence?
We also want to memoize the intermediate recursive calls, and in order
to achieve it, we need an explicit representation of the recursive
structure of the function.
\km{Should the encoding of recursive functions be motivated a little
  more?  explaining why the obvious
$(x0:dom -> (x:dom{x << x0} -> codom) -> codom)$ does not work ?}
Following \citet{McBride15}, we represent this by a function
\ls$x:dom -> partial_result x$, where a partial result is either a finished
computation of type \ls$codom$ or a request for a recursive call
together with a continuation.
\begin{lstlisting}
type partial_result (x0:dom) =
  | Done : codom -> partial_result x0
  | Need : x:dom{x << x0} -> cont:(codom -> partial_result x0) ->
         partial_result x0
\end{lstlisting}
As we define the fixed point using \ls$Need x f$, we crucially
require \ls$x << x0$, meaning that the value of the function is requested
at a point \ls$x$ where function's definition already exists.
For example encoding Fibonacci
amounts to the following code where the two recursive calls in the
second branch have been replaced by applications of the \ls$Need$
constructor.
We also define the fixpoint of such a function representation \ls$f$:
\begin{lstlisting}
let fib_skel (x:dom) : partial_result x =
  if x <= 1 then Done 1 else
    Need (x $-$ 1) (fun y$_1$ -> Need (x $-$ 2) (fun y$_2$ -> Done (y$_1$ + y$_2$)))
let rec fixp (f: x:dom -> partial_result x) (x0:dom) : codom =
  let rec complete_fixp x = function
    | Done y -> y
    | Need x' cont -> let y = fixp f x' in complete_fixp x (cont y)
  in complete_fixp x0 (f x0)
\end{lstlisting}
To obtain a memoized fixpoint, we need to memoize functions defined
only on part of the domain, \ls$x:dom{p x}$.
\begin{lstlisting}
let partial_memoize (p:dom -> Type) 
  (f : x:dom{p x} -> Memo codom) (x:dom{p x}) =
  match get x with Some y -> y | None -> let y = g x in put x y; y
let rec memoize_rec (f: x:dom -> partial_result x) (x0:dom) =
  let rec complete_memo_rec x :Memo codom = function
    | Done y -> y
    | Need x' cont -> 
      let y = partial_memoize (fun y -> y << x) (memoize_rec f) x' in 
      complete_memo_rec (cont y)
  in complete_memo_rec x0 (f x0)
\end{lstlisting}
\iffull Since both functions are syntactically similar i\else{I}\fi{}t
is relatively easy
to prove by structural induction on the code of \ls$memoize_rec$
that, for any skeleton of a recursive function
\ls$f$, we have that
%
\ls$(memoize_rec f) `computes` (fixp f)$.
\km{a slight complication is due to the decreasing clauses
but bringing them in will make the explanation pretty messy...}
The harder part is proving that \ls$fixp fib_skel$ is extensionally equal to
\ls$fibonacci$, the natural recursive definition of the sequence,
as these two functions are not syntactically similar---however, the 
proof involves reasoning only about pure functions.
As we have already proven that
\ls$memoize_rec fib_skel$ computes \ls$fixp fib_skel$, we easily gain a proof of the
equivalence of \ls$memoize_rec fib_skel$ to \ls$fibonacci$ by transitivity.
\iffull

Finally, we can encapsulate the \ls$Memo$ effect and provide a pure
state-passing interface:
\begin{lstlisting}
type memo_pack (f:dom -> codom) =
  | MemoPack : h0:memo_st{valid_memo h0 f} ->  
    mf:(dom -> Memo codom){mf `computes` f} -> memo_pack f
let apply_memo (#f:dom -> codom) (mp:memo_pack f) (x:dom) : 
    (codom * memo_pack f) =
  let MemoPack h0 mf = mp in let y, h1 = reify (mf x) h0 in 
  y, MemoPack h1 mf
let mk_memo_pack f : memo_pack (fixp f) = memo_lemma f ;
  MemoPack [] (memoize_rec f)
\end{lstlisting}
\fi
%

\subsection{Stepwise refinement and $n$-ary relations: Union-find with two optimizations}
\label{sec:unionfind}

In this section, we prove several classic optimizations of a
union-find data structure introduced in several stages, each a
refinement.
For each refinement step, we employ relational verification to prove
that the refinement preserves the canonical structure of union-find.
We specify correctness using, in some cases, $4$- and $6$-ary
relations, which are easily manipulated in our monadic framework.

\mypara{Basic union-find implementation} A union-find data structure
maintains disjoint partitions of a set, such that each
element belongs to exactly one of the partitions. The data structure
supports two operations: \ls$find$, that identifies to which partition an
element belongs, and \ls$union$, that takes as input two elements
and combines their partitions.

An efficient way to implement the union-find data structure is as a
forest of disjoint trees, one tree for each partition, where each node
maintains its parent and the root of each tree is the designated
representative of its partition.
The find operation returns the root of a given element's partition (by
traversing the parent links), and the union operation simply points
one of the roots to the other.

We represent a union-find of set
\ls@[0, n $-$ 1]@ as the type `\ls$uf_forest n$' (below),
a sequence of ref cells, where the \ls{i$^\textit{th}$} element in the
sequence is the \ls{i$^\textit{th}$} set element, containing its parent and the
list of all the nodes in the subtree rooted at that node. The list
is computationally irrelevant (\IE \emph{erased})---we only use it to
express the disjointness invariant and the termination metric for
recursive functions (e.g. \ls{find}).

\begin{lstlisting}
type elt (n:$\mathbb{N}$) = i:$\mathbb{N}${i $<$ n} $\times$ erased (list $\mathbb{N}$)
type uf$\_$forest (n:$\mathbb{N}$) = s:seq (ref (elt n)){length s = n}
\end{lstlisting}

\iffull
The liveness and disjointness invariants on a union-find forest are:

\begin{lstlisting}
(* all the refs are distinct and live in the heap *)
let live (#n:$\mathbb{N}$) (uf:uf$\_$forest n) (h:heap) =
($\forall$ i j. i $\not\eq$ j $\Rightarrow$ distinct uf[i] uf[j])  $\wedge$  ($\forall$ i. uf[i] $\in$ h)
let disjoint (#n:$\mathbb{N}$) (uf:uf$\_$forest n) (h:heap) =
  $\forall$ i. i $\in$ (subtree uf i h) $\wedge$  (* i is in its own subtree *)
     (* set$\_$n is the set of all numbers from 0 to n $-$ 1 *)
     subtree uf i h $\subseteq$ set$\_$n n $\wedge$
     (* i's subtree is a subset of its parent's subtree *)
     is$\_$root i $\vee$ subtree uf i h $\subset$ subtree uf (parent uf i h) h $\wedge$
     (* disjointness of subtrees *)
     $\forall$ j. (i $\not\eq$ j $\wedge$ is$\_$root uf i h $\wedge$ is$\_$root uf j h)
       $\Rightarrow$ subtree uf i h $\cap$ subtree uf j h = $\phi$
\end{lstlisting}

\fi

The basic \ls$find$ and \ls$union$ operations are shown below,
where \ls{set} and \ls{get} are stateful functions that read and
write the \ls{i$^\text{th}$} index in the \ls{uf} sequence.
Reasoning about mutable pointer structures requires maintaining
invariants regarding the liveness and separation of the memory
referenced by the pointers. While important, these are orthogonal to
the relational refinement proofs---so we elide them here,
but still prove them intrinsically in our code.

\begin{lstlisting}
let rec find #n uf i = let p, $\_$ = 
  get uf i in if p = i then i else find uf p 
let union #n uf i$_1$ i$_2$ = let r$_1$, r$_2$ = find uf i$_1$, find uf i$_2$ in 
    let $\_$, s$_1$ = get uf r$_1$ in let $\_$, s$_2$ = get uf r$_2$ in
    if r$_1$ <> r$_2$ then (set uf r$_1$ (r$_2$, s$_1$); set uf r$_2$ (r$_2$, union s$_1$ s$_2$)) 
\end{lstlisting}

\mypara{Union by rank} The first optimization we consider
is \iffull improving \ls$union$ to \fi \ls$union_by_rank$, which decides whether
to merge \ls@r$_1$@ into \ls@r$_2$@, or vice versa, depending on the
heights of each tree, aiming to keep the trees shallow.
We prove this optimization in two steps, first refining the
representation of elements by adding a rank field to \ls$elt n$ and then
proving that \ls$union_by_rank$ maintains the same set partitioning
as \ls$union$.

\begin{lstlisting}
type elt (n:$\mathbb{N}$) = i:$\mathbb{N}${i $<$ n} $\times$ $\mathbb{N}$ $\times$ erased (list nat) (* added rank *)
\end{lstlisting}

We formally reason about the refinement by proving that the outputs of
the \ls{find} and \ls{union} functions do not depend on the newly
added rank field.
The \ls$rank_independence$ lemma (a $4$-ary relation) states
that \ls{find} and \ls{union} when run on two heaps that differ only
on the rank field, output equal results and the resulting heaps also
differ only on the rank field.
\begin{lstlisting}
let equal_but_rank uf h$_1$ h$_2$ =  $\forall$ i. parent uf i h$_1$ = parent uf i h$_2$
                             $\wedge$ subtree uf i h$_1$ = subtree uf i h$_2$
let rank$\_$independence #n uf i i$_1$ i$_2$ h$_1$ h$_2$ : Lemma
(requires (equal$\_$but$\_$rank uf h$_1$ h$_2$))
(ensures  (let (r$_1$,f$_1$), (r$_2$,f$_2$) = 
  reify (find uf i) h$_1$,reify (find uf i) h$_2$ in
 let ($\_$,u$_1$), ($\_$,u$_2$) = 
  reify (union uf i$_1$ i$_2$) h$_1$,reify (union uf i$_1$ i$_2$) h$_2$ in
 r$_1$ == r$_2$ $\wedge$ equal$\_$but$\_$rank uf f$_1$ f$_2$ /\ equal$\_$but$\_$rank uf u$_1$ u$_2$))
\end{lstlisting}

\iffull

\mypara{Union by rank} The rank based union optimization aims at
minimizing the height of the subtrees, so that the tree traversal is
more efficient. It does so by pointing the root with smaller height to
the other root during the union operation.

\begin{lstlisting}
let union$\_$opt #n uf i$_1$ i$_2$ =
  let r$_1$, r$_2$ = find uf i$_1$, find uf i$_2$ in
  let $\_$, d$_1$, s$_1$ = get uf r$_1$ in let $\_$, d$_2$, s$_2$ = get uf r$_2$ in
  if r$_1$ = r$_2$ then ()
  else begin
    if d$_1$ $<$ d$_2$ then begin (* point r$_1$ to r$_2$ *)
      set uf r$_1$ (r$_2$, d$_1$, s$_1$); set uf r$_2$ (r$_2$, d$_2$, union s$_1$ s$_2$)
    end
    else begin  (* point $r_2$ to r$_1$ and adjust $r_1$'s height *)
      set uf r$_2$ (r$_1$, d$_2$, s$_2$);
      let d$_1$ = if d$_1$ = d$_2$ then d$_1$ + 1 else d$_1$ in
      set uf r$_1$ (r$_1$, d$_1$, union s$_1$ s$_2$)
    end
  end
\end{lstlisting}
\fi

Next, we prove the \ls$union_by_rank$ refinement
sound. Suppose we run \ls{union} and \ls$union_by_rank$ in \ls{h}
on a heap \ls{h} producing \ls{h$_1$} and \ls{h$_2$}.
Clearly, we cannot prove that find for a node \ls{j} returns the same
result in \ls{h$_1$} and \ls{h$_2$}. But we prove that the canonical
structure of the forest is the same in \ls{h$_1$} and \ls{h$_2$}, by
showing that two nodes are in the same partition in \ls{h$_1$} if and only if
they are in the same partition in \ls{h$_2$}:

\begin{lstlisting}
val union_by_rank_refinement #n uf i$_1$ i$_2$ h j$_1$ j$_2$ : Lemma
  (let (_, h$_1$), (_, h$_2$) = 
    reify (union uf i$_1$ i$_2$) h, reify (union_by_rank uf i$_1$ i$_2$) h in
   fst (reify (find uf j$_1$) h$_1$) == fst (reify (find uf j$_2$) h$_1$) <==> 
     fst (reify (find uf j$_1$) h$_2$) == fst (reify (find uf j$_2$) h$_2$))
\end{lstlisting}

This property is $6$-ary relation, relating 1 run of \ls$union$ and
1 run of \ls$union_by_rank$ to 4 runs of \ls$find$---its proof is a
relatively straightforward case analysis.

\mypara{Path compression} Finally, we consider \ls$find_compress$,
which, in addition to returning the root for an element, sets the root
as the element's new parent to accelerate subsequent find queries.
\iffull

\begin{lstlisting}
let rec find$\_$opt #n uf i =
  let p, d, s = get uf i in
  if p = i then i
  else
    let r = find$\_$opt uf p in
    set uf i (r, d, s);
    r
\end{lstlisting}

\fi
To prove the refinement of \ls{find} to \ls{find_compress} sound, we
prove a $4$-ary relation showing that if running \ls{find} and
\ls{find_compress} on a heap \ls{h} results in the heaps \ls{h$_1$} and \ls{h$_2$},
then the partition of a node \ls{j} is the same in \ls{h$_1$}
and \ls{h$_2$}. This also implies that \ls{find_compress} retains the
canonical structure of the union-find forest.

\begin{lstlisting}
val find_compress_refinement #n uf i h j
  : Lemma (let (r$_1$, h$_1$), (r$_2$, h$_2$) = 
    reify (find uf i) h, reify (find_compress uf i) h in
    r$_1$ == r$_2$ $\wedge$ fst (reify (find uf j) h$_1$) == fst (reify (find uf j) h$_2$))
\end{lstlisting}

\section{Related work}
\label{sec:related}

\ch{If it becomes public in time also cite: Modular Product
  Programs. Marco Eilers, Peter Müller, and Samuel Hitz. Submitted to ESOP'17.}



Much of the prior related work focused on checking specific relational
properties of programs, or general relational properties using
special-purpose logics.
In contrast, we argue that proof assistants that support reasoning
about pure and effectful programs can, using our methodology, model
and verify relational properties in a generic way.
The specific incarnation of our methodology in \fstar exploits its
efficient implementation of effects enabled by abstraction and
controlled reification; a unary weakest precondition calculus as a
base for relational proofs; SMT-based automation; and the convenience
of writing effectful code in direct style with returns, binds, and
lifts automatically inserted.


\mypara{Static IFC tools}
%
\citet{SabelfeldM03} survey a number of IFC type systems and
static analyses for showing noninterference, trading completeness for
automation.
More recent verification techniques for IFC aim for better
completeness \cite{BeringerH07, NanevskiBG13, AmtoftB04, AmtoftDZABHOC12,
BanerjeeNN16, SchebenS11, BartheFGSSB14, Rabe16}, while compromising
automation.
The two approaches can be combined, as discussed in
in \autoref{sec:ifc-combining}.

\mypara{Relational program logics and type systems}
A variety of program logics for reasoning about general relational
properties have been proposed previously \citep{benton04relational,
Yang07, BartheGB09, AguirreBGGS17}, while others apply general
relational logics to specific domains, including access
control \cite{NanevskiBG13}, cryptography \cite{BartheGB12,
BartheGB09, BartheDGKSS13, PetcherM15}, differential
privacy \cite{BartheKOB13, ZhangK17}, mechanism
design \cite{BartheGAHRS15}, cost analysis \cite{CicekBG0H17}, program
approximations \cite{CarbinKMR12}.

R\fstar, is worth pointing out for its
connection to \fstar. \citepos{BartheFGSSB14} extend a prior, value-dependent
version of \fstar \cite{fstar-pldi13} with a probabilistic semantics
and a type system that combines pRHL with refinement types. Like many
other relational Hoare logics, R\fstar provided an incomplete set of
rules aimed at capturing many relational properties by intrinsic typing only.

In this paper we instead provide a versatile generic method for relational
verification based on modeling effectful computations using monads and
proving relational properties on their monadic representations, making
the most of the support for full dependent types
and SMT-based automation in the latest version of \fstar.
This generic method can both be used directly to verify programs or as
a base for encoding specialized relational program logics.

\ch{Can we encode R\fstar?}



\mypara{Product program constructions}
%
Product program constructions and self-composition are techniques
aimed at reducing the verification of k-safety properties
\cite{clarkson10hyp} to the
verification of traditional (unary) safety proprieties of a product
program that emulates the behavior of multiple input programs.
Multiple such constructions have been proposed \cite{BartheCK16} targeted for
instance at secure IFC \cite{TerauchiA05, BartheDR11, Naumann06, YasuokaT14},
program equivalence for compiler validation \cite{ZaksP08},
equivalence checking and computing semantic differences
\cite{LahiriHKR12}, program approximation \cite{HeLR16}.
\citepos{SousaD16} recent Descartes tool for k-safety properties also
creates k copies of the program, but uses lockstep reasoning to
improve performance by more tightly coupling the key invariants across
the program copies.
Recently \citet{AntonopoulosGHKTW17} propose a tool \iffull called Blazer \fi that
obtains better scalability by using a new decomposition of
programs instead of using self-composition for k-safety problems.
\ch{So how do we relate to all this? What we do is quite
  different. Our approaches at this like compose2 didn't work so great
  (see plan.org). We have better support for interaction? Better
  expressiveness? Higher-order programs?
  Nik discussed with the authors of \cite{SousaD16} before ICFP?}
\ch{Should still figure this out since Terauchi and Dillig
  are likely to review this}
\ch{They can't do loop reversal, quite syntactic}



\mypara{Other program equivalence techniques}
Beyond the ones already mentioned above, many other techniques targeted at program
equivalence have been proposed; we briefly review several recent works:
\citet{BentonKBH09} do manual proofs of correctness of
compiler optimizations using partial equivalence relations.
\citet{KunduTL09} do automatic translation validation of compiler
optimizations by checking equivalence of partially specified programs
that can represent multiple concrete programs.
\citet{GodlinS10} propose proof rules for proving the
equivalence of recursive procedures.
\citet{LucanuR15} and \citet{CiobacaLRR16} generalize this to
a set of co-inductive equivalence proof rules that are language-independent.
Automatically checking the equivalence of processes in a process
calculus is an important building block for security protocol
analysis \cite{BlanchetAF08, ChadhaCCK16}.

\ch{Again, how do we relate to all this?}


\mypara{Semantic techniques}
Many semantic techniques have been proposed for reasoning about
relational properties such as observational equivalence, including
techniques based on binary logical relations \cite{BentonKBH09,
  Mitchell86, AhmedDR09, DreyerNRB10, DreyerAB11, DreyerNB12,
  BentonHN13, Benton0N14}, bisimulations \cite{KoutavasW06,
  SangiorgiKS11, Sumii09} and combinations thereof \cite{HurDNV12, HurNDBV14}.
While these very powerful techniques are often not directly automated,
they can be used to provide semantic correctness proofs for relational
program logics \cite{DreyerNRB10, DreyerAB11} and other verification
tools \cite{BentonK0N16}.

\iffull
\section{Future work}
\label{sec:discussion}



While we found \fstar to be a versatile tool for relational
verification of effectful programs, we also contemplated about
features that would make it even better suited.

\mypara{Tactics} \fstar's current combination of SMT
solving and dependent typechecking \iffull with higher-order unification and
normalization \fi provides good automation, but the ongoing addition of tactics
will provide more control and the possibility of
user-defined decision procedures.
In particular, when using shallow embeddings (like we do in
\autoref{sec:transformations}) tactics will allow us to write
meta-programs that automatically apply derived proof rules
based on the structure of the \fstar{} program we want to verify.



\mypara{Extrinsic termination reasoning}
Aside from their use in relational reasoning, extrinsic proofs of
reified terms allow programmers to defer proof obligations, rather
than insisting on proofs at the time of
definition (while anticipating all uses).
While convenient, extrinsic proofs in \fstar only apply to programs
that are intrinsically proved terminating.
Building on our use of \citepos{McBride15} approach
in \S\ref{sec:memo}, we aim to define divergence as a reifiable
effect, placing it on par with other effects in \fstar.
We could then reason about the partial correctness of a program
declared in this effect or to prove its termination after its
definition.
Going back to the \emph{while} interpreter from
\autoref{sec:ifc-while}, we could forget about the decreasing metric and use
either \citepos{BoveC05} termination witnesses or step-indexing
as in \autoref{sec:benton2004-rhl} \cite{OwensMKT16,AminR17},
proving, for example, noninterference of reachable states
of an interactive non-terminating program.

\iflater
\km{There are also nice ideas in papers by Andreas Abel and others about
  productive coprogramming and verification of such code but It may feel
as totally unrelated to anything F* currently has}
\fi

\mypara{Observational purity}
%
Another desirable feature would be to hide the effect of a term if it
is proven observationally pure, e.g., in \autoref{sec:memo} this would
provide the ability to replace the original pure code by its
equivalent memoized variant.
Since we are able to prove that the memoized code has the same
extensional behaviour as the pure code up to some private data that we
could abstract over, we would like to implement a mechanism to
encapsulate observationally pure code.
%
We hope that this mechanism could also be applied to programs proven
terminating extrinsically.


\km{Is there any valid reference to Danel's monotonicity?}
\ch{\cite{preorders}, not clear what the relevance is}

\ch{missing references on forgetting allocation effects: \cite{BentonHN13, Benton0N14}}

\ch{One more citation that could be relevant for the While language
  (although this one is based on coinduction):
  Operational semantics using the partiality monad.
  \url{http://www.cse.chalmers.se/~nad/publications/danielsson-semantics-partiality-monad.pdf}
}

\fi

\section{Conclusion}

This paper advocates verifying relational properties of effectful
programs using generic tools that are not specific to relational
reasoning: monadic effects, reification, dependent types,
non-relational weakest preconditions, and SMT-based automation.
Our experiments in \fstar{} verifying relational properties about a
variety of examples show the wide applicability of this approach.
One of the strong points is the great flexibility in modelling effects
and expressing relational properties about code using these effects.
The other strong point is the good balance between interactive control,
SMT-based automation, and the ability to encode even more automated
specialized tools where needed.
Thanks to this, the effort required from the \fstar{} programmer for
relational verification seems
on par with non-relational reasoning in \fstar{} and with specialized
relational program logics.




\begin{acks}                            

The work of C\u{a}t\u{a}lin Hri\c{t}cu and Kenji Maillard is
in part supported by the
\grantsponsor{1}{European Research Council}{https://erc.europa.eu/}
under ERC Starting Grant SECOMP (\grantnum{1}{715753}).
\end{acks}


\ifcamera
\balance
\clearpage
\bibliographystyle{ACM-Reference-Format}
\else
\bibliographystyle{abbrvnaturl}
\fi
\bibliography{fstar}

\end{document}